# Ion Sources for Production of Highly Charged Ion Beams

Snowmass2021 Ion Source Working Group

Lead Editor: A. Lapierre[1]
Co-authors: J. Benitez[2], A. Lapierre[1], M. Okamura[3], D. Todd[2], D. Xie[2]
Convener: Y. Sun[4]

[1] Facility for Rare Isotope Beams, Michigan State University, 640 Shaw Ln, E. Lansing, MI 48824
[2] Berkeley National Laboratory, 1 Cyclotron Rd, Berkeley, CA 94720
[3] Brookhaven National Laboratory, 98 Rochester St, Upton, NY 11973
[4] Argonne National Laboratory, 9700 S Cass Ave, Lemont, IL 60439

## 1. Introduction

### 1.1. General motivation

Intense Highly Charged Ion Beams (HCIB) from injector ion sources at heavy ion accelerator facilities, such as the Electron-Ion Collider (EIC) [EIC2022] and the proposed Future Circular Collider (FCC) [FCC2022] are in demand to expand research in particle physics. Future facilities, like the High-Intensity Heavy Ion Facility [HIAF2022], and future upgrades to existing ones to enhance their capabilities, such as the Facility for Rare-Isotope Beams (FRIB) [FRIB2022], also demand higher charge states and HCIB of higher intensities for production of secondary beams needed in nuclear physics research. In addition to particle and nuclear physics research, HCIB are also needed for the radiation effects testing community as the growing complexity of microelectronics requires ions of higher energies.

At Rare-Isotope Beam (RIB) facilities, ion sources used as charge breeders of rare (radioactive) isotopes convert ion beams (typically 1+) injected from a production target into HCIB for (re-)acceleration with a post-accelerator. Rare isotopes are produced in small quantities and can have short decay times ($\lesssim$100 ms). Efficient and fast breeding of HCIB that are free of contaminants is of uttermost importance. With present and future upgrades to RIB facilities, the production rates are expected to significantly increase. Charge-breeding systems capable of handling intense injected beams are needed, without compromising efficiency and beam purity.

With accelerator upgrades and advances in accelerator technologies, not all HCIB demands can be met with existing injector and charge-breeder ion sources. Continued research and development (R&D) in this field are therefore essential to continually improve their performance and match the unprecedented and increasingly higher HCIB requirements from the accelerator community.

This White Paper discusses the present production capabilities of ion sources of HCIB, and the potentials of future highly charged ion (HCI) sources. It also discusses the strengths and weaknesses of such sources, along with paths forward for improving their performance to meet the requirements of present and future heavy ion accelerator facilities. Advanced computer science tools such as Artificial Intelligence and Machine Learning, also discussed here, can be used to increase our understanding and enhance the production of these advanced HCIB sources. This document is meant to be utilized as a basis to guide the conceptual design of future heavy ion accelerators.

**1.2. Introduction to HCI production, beam properties, and ion sources**

In intense HCI sources, the ions are primarily created by sequential single electron Coulomb collisions. The (single) electron-impact ionization cross section depends on the ionization potential (electron binding energy) of the ions or atoms being ionized, details of their electronic structures, and the energy of the free electrons colliding with the ions [BEYE2002]. This cross section decreases for higher charge states as the ionization potential increases, sometimes rather dramatically. As shown in Fig. 1.2.1, a difference of almost six orders of magnitude is clearly seen between the cross sections for ionization to singly ionized and fully stripped argon ions, demonstrating the difficulties in generating the bare ions.

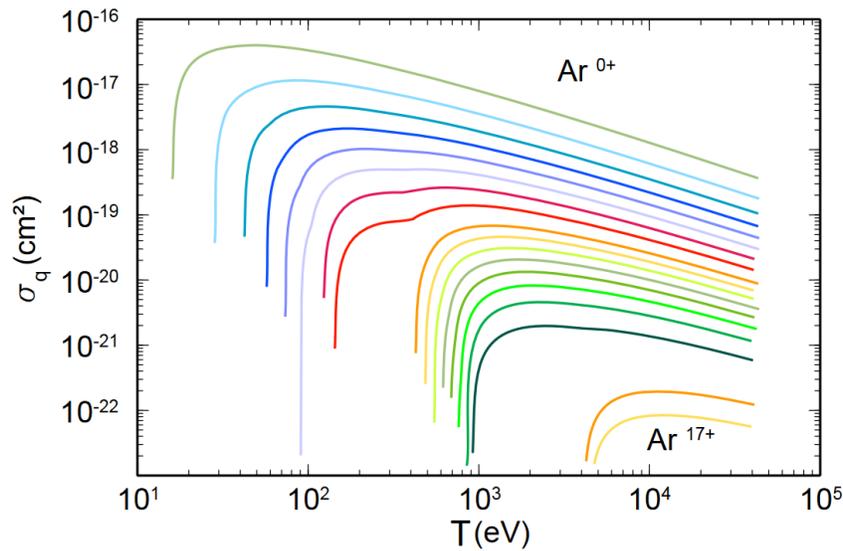

*Figure 1.2.1. Electron-impact ionization cross-section of the argon atom and ions in dependence of electron energy T, (taken from [ZSCH2013]).*

In addition to single electron-impact ionization and weaker high orders, other dominating atomic-physics processes contribute to modifying the distribution of charge states in ion sources such as radiative and resonant multi-electronic recombination with (free) electrons as well as charge-exchange recombination with neutral (residual or injected) gases. Figure 1.2.2 shows how the cross section of charge-exchange, electron-impact ionization, and electron recombination exponentially varies for krypton charge states. A high electron density is essential for the ionization rate to exceed the charge-exchange rate with neutral residual and/or injected gases, which significantly increases with the ion charge state. This is often a limiting factor for production of HCI in ion sources.

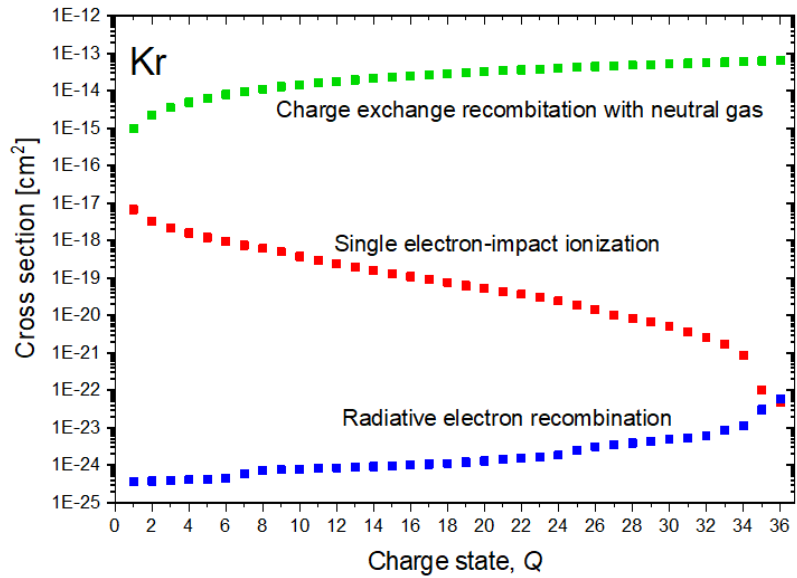

*Figure 1.2.2. Cross section of (single) electron-impact ionization compared with that of competing processes: charge-exchange with residual or injected neutral gases and (radiative) recombination with free electrons.*

Optimum production of HCI, which have substantially smaller ionization cross sections, requires a sufficiently high electron density $n_e$, characteristic ionization time $\tau_i$, and electron energy (T). The product $n_e\tau_i$ can be seen as an ionization factor; a figure of merit that is a fundamental quantity that can describe the performance of an HCI source. Figure 1.2.3 schematically summarizes the relationship of these parameters for generating a desired ion charge state, *Q*. It shows the optimum values of the ionization factor and electron energy for selected elements, considering their electron-impact ionization cross-sections.

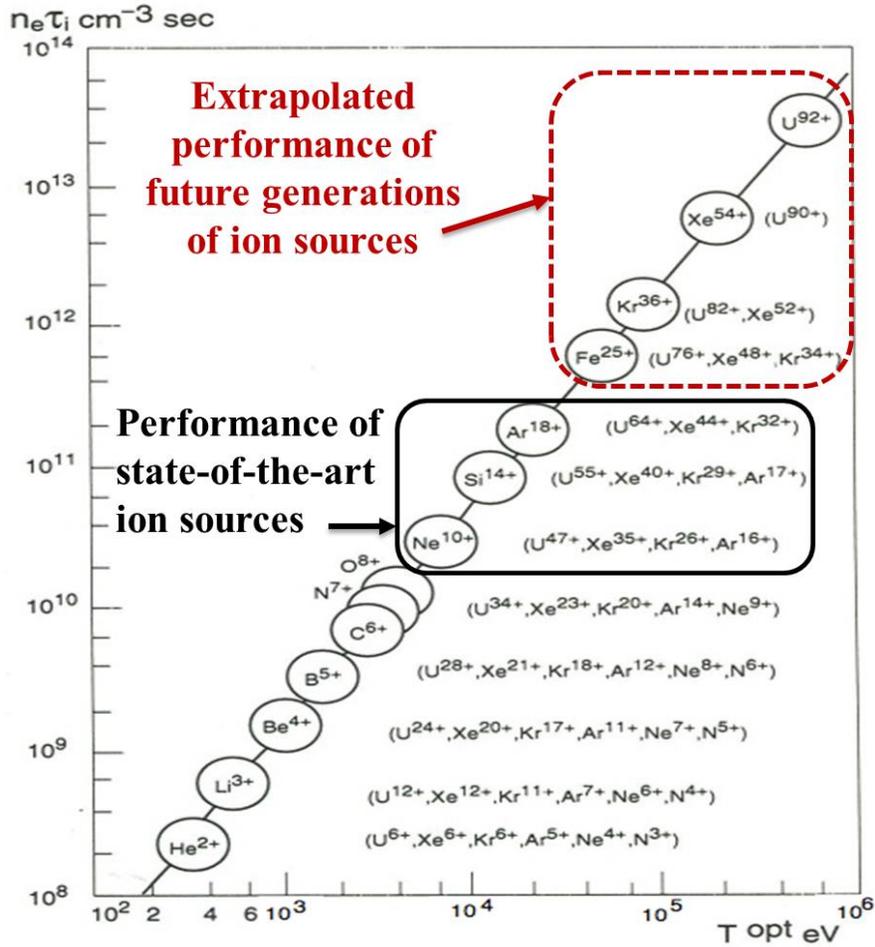

*Figure 1.2.3. The theoretical relationship of the product ($n_e\tau_i$) of the electron density and ionization time with the electron energy (T) needed for generating HCIB of ion charge Q. The solid rectangular circle indicates the highest capabilities of present state-of-the-art ion sources and the extrapolated performance (indicated by a red dashed rectangular circle) of the future generations of ion sources.*

HCI sources at heavy ion accelerator facilities are predominantly used to extend the energy range of the accelerated beams. The gain in ion beam energy strongly depends on the accelerating electrical field and charge state of the accelerated ions, either proportional to $Q^2$ for a cyclotron or $Q$ for a linear accelerator (LINAC). For the same energy gain, less accelerating field is then needed with a higher $Q$ and similarly, a higher ion beam energy gain can be achieved for the same accelerating field. Availability of ion beams of increasingly higher $Q$ and lower atomic mass-over-charge ratio (A/Q) allows the energy range of accelerators to be extended at a lower cost while reducing operating constraints on facilities.

The beam intensity *I*, transverse emittance *ε*, and energy spread *ΔE* are important quantities defining the quality of ion beams for accelerators, in addition to its *A/Q*. HCI sources have to provide beams of sufficient intensity as well as of small emittance and energy spread, generally accepted to yield the best results for optimum transport (efficiency), designs, and operations of heavy ion accelerators. Another important quantity is the time structure of the extracted ions from the sources that can be either pulsed or continuous to match accelerators' capabilities and users' requirements.

Electron Cyclotron Resonance Ion Sources (ECRIS), Electron-Beam Ion Sources (EBIS), and Laser Ion Sources (LIS) provide HCIB of best quality that meet the requirements of accelerators. They are predominantly employed at heavy ion accelerator facilities for generating intense HCIB. Charge-breeder ion sources based on ECRIS and EBIS are used at post-accelerator RIB facilities for their high efficiencies and short breeding times. Table 1.2.1 briefly lists typical parameters of HCIB obtained from these ion sources, which are described in more detail in the following sections. Due to an uncontrollable circumstance, the description of the EBIS was unavailable at the submission time of the White Paper and thus had to be excluded.

*Table 1.2.1. Present typical performance and beam qualities of ion sources for HCIB.*

| | ECRIS | LIS | Charge Breeder | |
| --- | --- | --- | --- | --- |
| | | | EBIST | ECRIS |
| **Beam variety** | All-natural elements & isotopes | Except noble gas [d] | Radioactive isotopes | |
| $Q_{max}$ [a, e] | ~$U^{64+}$ | $C^{6+}$, $Fe^{26+}$ | ~$U^{64+}$ | ~$Rb^{20+}$ |
| $A$ (u) | 2 - 238 | ≥ 3 (up to clusters) | < 200 (achieved so far for experiments) | |
| A/Q | 2-7 | 2-7 | 2-7 | |
| $\varepsilon$ [b] (mm mrad) | ~ 0.2 | 0.05 - 1 | ≲0.04 | ≲0.08 |
| $\Delta E$ (eV) | ~ 20 - 50 | 1 - 1k | ≲300 | ≲50 |
| Time structure | Continuous | Pulsed | Pulsed | Continuous |
| **Intensity** (ions/sec) | ~ $10^{14}$ – $10^{16}$ [c] | $10^{10}$ ~ $10^{13}$ per laser pulse | ≲$10^6$ | |

[a] Unit of elementary charge (1 charge = 1.602×$10^{-19}$ C).
[b] Root-Mean-Square (RMS) normalized emittance.
[c] Maximum beam intensity for A/Q = 2-7.
[d] Frozen noble gas can be used as a laser target.
[e] ~$U^{64+}$ is estimated based an achievable breeding times (~ 1-2 s); ~$Rb^{20+}$ is observed.

## 2. Ion sources for HCIB

This section reviews the basic design, state-of-the-art operation principle, and present performance of ion sources for HCIB production at heavy ion accelerator facilities. It also reviews the R&D needs to improve the performance of these sources and discusses their capabilities for future accelerator developments.

### 2.1. Electron Cyclotron Resonance Ion Source (Lead Contributor: **Xie**)

A high charge state Electron Cyclotron Resonance Ion Source (ECRIS) can produce intense HCIB in continuous wave (cw) or pulsed mode from a plasma confined in magnetic fields. It is a spin-off from fusion plasma research using microwaves to energize a plasma for the generation of ions. The development of successive generations of ECRIS, following the first in the early 1970s, have followed the path summarized by the inventor Geller in 1987:

$$I \propto n_c \propto \omega^2 \propto B_{ecr}^2$$

This is the basic ECRIS "scaling law" which predicts the generated current $I$ of the highly charged $Q$ ion beam is proportional to the square of the microwave heating frequency (cut-off plasma density, $n_c \propto \omega^2$) or the square of the magnetic resonance field $B_{ecr}$ ($n_c$ is also $\propto B_{ecr}^2$) [GELL1987].

#### 2.1.1. Basics of an ECR Ion Source

An ECRIS uses a minimum-$B$ field configuration (the most critical source component) with high-strength magnetic mirrors to provide fields for resonant electron heating and for overall plasma confinement. Microwaves of a few frequencies $\omega$, multiple-frequency heating, are chosen to efficiently produce energetic electrons via Electron Cyclotron Resonance Heating on a number of closed resonance surfaces within the minimum-$B$ field. These energetic electrons ionize neutrals and ions within the source via successive collisions. The typical ion energy of charge state $Q$ in the plasma is a few $Q$ eV, while the electron energy distribution ranges from low eV to above 1 MeV. The high end of the electron energy is sufficient to ionize hydrogen-like uranium ions (~ 130 keV binding energy).

As schematically indicated in Figure 2.1.1a, the typical minimum-$B$ field for an ECRIS is a superposition of an axial mirror field produced by a set of solenoids, as shown in Figure 2.1.1b, and a pure radial field proportional to the square of radius that is produced by a set of sextupole racetrack coils, as shown in Figure 2.1.1c. ECRIS development has shown that the ECR plasma is insensitive to the specifics of the magnet coil geometry as long as a high-strength, minimum-$B$ field is produced. Recent studies with VENUS [LYNE1997] at the Lawrence Berkeley National Laboratory (LBNL) have demonstrated that a cylindrical plasma chamber is not a necessity for a high charge state ECR ion source, as the microwaves of tens of GHz result in centimeter and millimeter wavelengths can easily travel into and resonate in a large complex-shaped plasma chamber of 3-5 Liter in volume. An asymmetry to the ECR plasma chamber results in greater flexibility in chamber design for the development of a high charge state ECRIS.

ECRIS can produce ions of all naturally existing elements as long as the material can be introduced into the plasma. Depending on its physical and chemical properties and ease of operations, introduction of the working substance to an ECRIS plasma is through gas valves, vaporizing ovens, or sputtering materials. With such a great variety of ions available from its ECRIS, the 60-year aged 88-Inch Cyclotron at LBNL has so far accelerated 49 elements and many isotopes, as shown in Fig. 2.1.2. This is a record for any heavy ion accelerator worldwide.

Because of its remarkable capability, versatility and, reliability, the ECRIS has been widely used for the production of intense highly charged ion beams for accelerators worldwide since the 1980s.

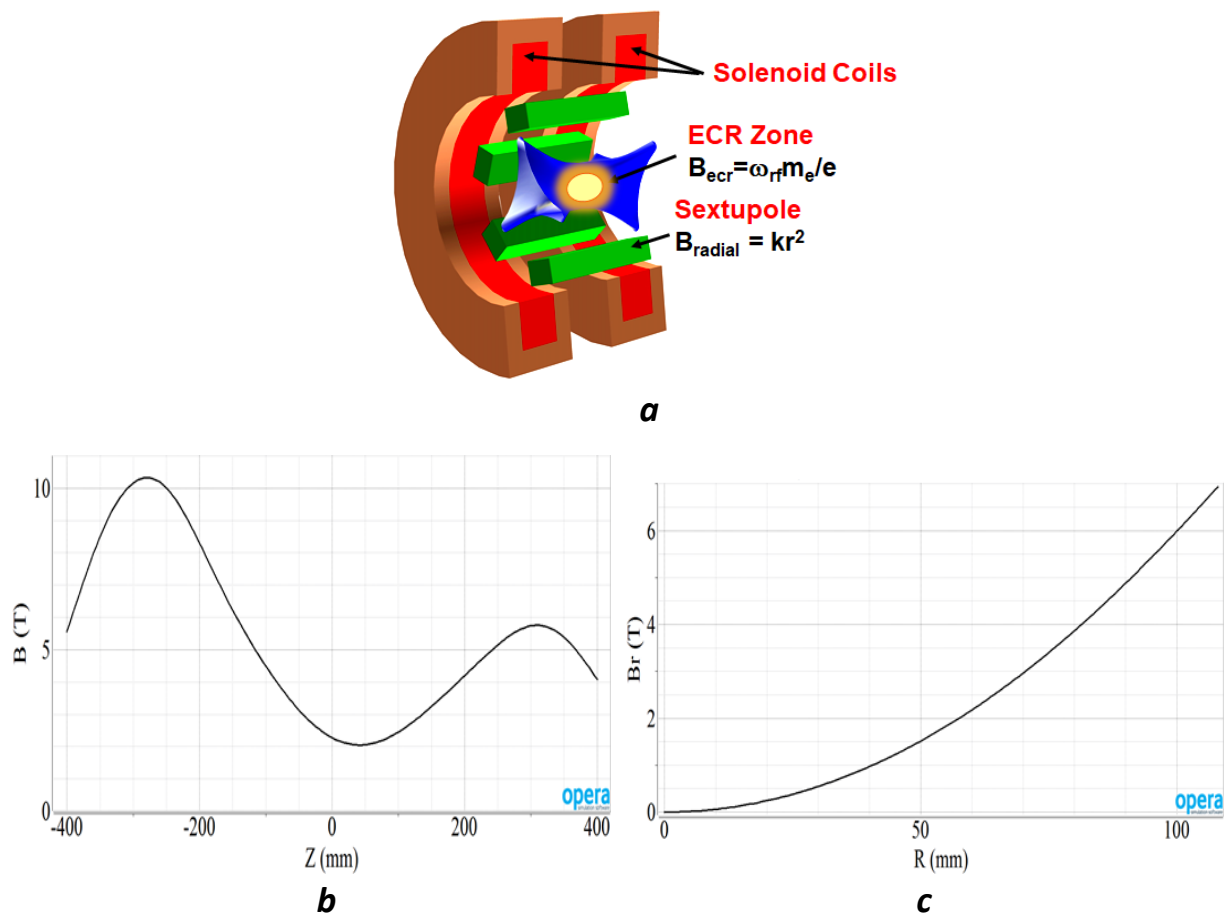

Figure 2.1.1. *a*. A schematic view of the magnetic structure of an ECRIS minimum-B field. *b*. The axial mirror field profile produced by a set of solenoids. *c*. The radial field profile produced with a sextupole (proportional to the square of the radius R).

*Figure 2.1.2. Since its commencement in 1962, the 88-Inch Cyclotron at LBNL has accelerated 49 elements produced by ECRIS (indicated with blue color squares), which is more than one-half of the natural elements, and a great number of isotopes.*

### 2.1.2. The current state-of-the-art ECR Ion Source

The 3rd generation ECRIS is the current state-of-the-art for these sources and typically consists of the source components listed below and shown in Figure 2.1.3:

- A minimum-*B* field (produced by a superconducting NbTi magnet housed in a cryostat),
- A plasma chamber (a cavity that houses the hot ECR plasma),
- A set of microwave generators (multiple-frequency plasma heating),
- An injection assembly (for gas input and for microwave coupling into the plasma chamber),
- An extraction mechanism (for ion beam extraction),
- An ion beam analysis system (typically a dipole magnet for selecting ion species),
- A beam transport line,
- System vacuum pumps.

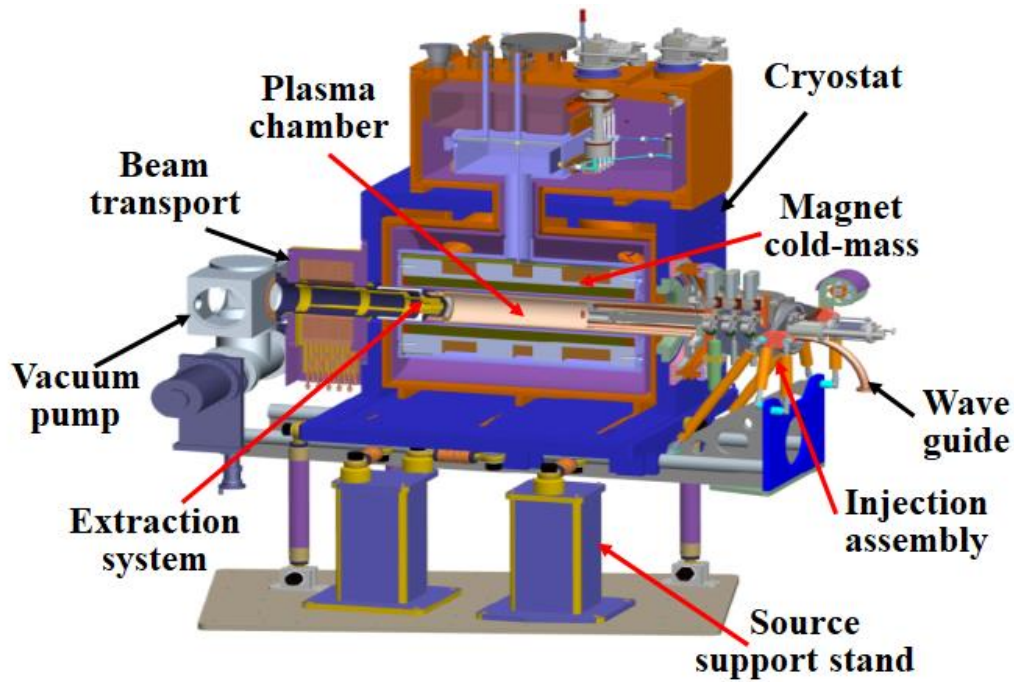

*Figure 2.1.3. This figure shows most of the components of a 3rd generation ECRIS using a superconducting magnet for the generation of a high-strength minimum-B field. The microwave generators and the beam analysis magnet are not shown.*

As Geller's scaling law predicts, the 3rd generation ECRIS, with field maxima of 4 T axially and 2.2 T radially, have substantially advanced the performance of ECRIS. Typical beam intensities of present 3rd generation ECRIS and their comparisons to the previous generations operating at lower magnetic fields and lower frequencies are listed in a combined-comparison table (Table 2.1.2). Listed below are the present 3rd generation ECRIS in operation or under development worldwide with maximum heating microwave power of about 10 – 15 kW:

- USA. Three:
    - A 28 GHz VENUS/LBNL [LYNE1997] and a
    - 24 GHz SUSI/NSCL/MSU, and a 28 GHz VENUS-II/FRIB/MSU [ZAVO2008], [REN2020)
- Japan. One: 28 GHz SC-ECR/RIKEN [NAKA2010]
- China. Two:
    - A 24 GHz SECRAL-I [ZHAO2006] and a
    - 28 GHz SECRAL-II/IMP [SUN2016]
- South Korea. Two under development:
    - A 28 GHz SC ECR for RAON/RISP [PARK2016] and a
    - 28 GHz SC ECR for a KBSI accelerator-based fast neutron source. [HEO2016]

### 2.1.3. Future generations of ECR Ion Sources

Future heavy ion accelerators can benefit greatly from the increased performance of a next-generation ECRIS operating at higher magnetic fields and higher frequencies. The 4th generation of ECRIS will operate at frequencies of ~45 GHz, almost doubling the operation frequency of 28 GHz currently used in 3rd generation ECRIS. The critical hurdle for a 4th generation ECRIS is to generate the magnetic fields needed for providing better confinement of the higher temperature plasma. The required magnetic fields, based on empirical design criteria, are listed in Table 2.2.1 for future operations of 45 – 84 GHz and compared to that of 28 GHz. A 45 GHz ECRIS should lead to a 2.5-fold increase in plasma densities and correspondingly larger ion beam intensities. The resulting charge state distributions for intense heavy ion beams will facilitate achieving the accelerator performance requirements of new facilities, possibly leading to substantial cost savings.

Table 2.1.1. Magnetic fields needed for future generations of ECR ion sources operating at 45-84 GHz and comparison to VENUS at 28 GHz.

| Magnetic Fields | Future Gen of ECRIS $\omega$ (45 – 84 GHz) | 3rd Gen VENUS $\omega$ (28 GHz) |
|---|---|---|
| $B_{ecr}$ (T) | 1.6 to 3 | 1 |
| $B_{Inj}$ (T) | ~ 5.6 to 10.5 (3.5-4 · $B_{ecr}$) | 4 |
| $B_{Min}$ (T) | ~ 0.8 to 2.4 (0.5 to 0.8 · $B_{ecr}$) | 0.5-0.8 |
| $B_{Ext}$ (T) | ~ 3.2 to 6 (~ 2 · $B_{ecr}$) | 2.2 |
| $B_{Rad}$ (T) | ~ 3.2 to 6 (2 · $B_{ecr}$)* | 2.0* |

*Radial field strength at the plasma chamber walls.*

All 3rd generation ECRIS have been built with NbTi superconducting magnets using traditional racetrack coils for the sextupole and operate at the conductor constraint limited by its upper critical field of about 10 T at 4.2 K. Going from a 3rd generation to a 4th generation 45 GHz ECRIS, the magnetic fields should be increased by about a factor of 1.6 that would lead to the same increase on the coil currents of the magnet built with the existing magnet geometries. This current increase would necessitate the use of a new, higher-current conductor such as $Nb_3Sn$ that can provide sufficient current density for higher field productions. However, $Nb_3Sn$ has a low yield stress of ~ 250 MPa, in comparison to ~ 400 MPa for the NbTi conductor, and much lower quench propagation that could result in large-size coils with serious engineering difficulties. Since a Nb$_3$Sn magnet has never been built and tested for an ECRIS, it would be essential to address all the uncertainties prior to building such a magnet.

Presently, there are two ongoing developments for a 45 GHz ECRIS. One is conducted by the ion source group of the Institute of Modern Physics (IMP) in Lanzhou/China and the other is carried out by the ECRIS group at the 88-Inch Cyclotron Facility of LBNL, USA. These developments are discussed below.

*2.1.3.1. Developing a high field Nb$_3$Sn Sextupole-in-Solenoid magnet*

The ECRIS group of the Institute of Modern Physics (IMP) in Lanzhou/China is the only group in the world currently attempting to construct a Nb$_3$Sn magnet for a 45 GHz ECRIS. [ZHAO2018] As shown in Figure 2.1.4, the Nb$_3$Sn magnet for their source (named FECR) is built using a geometry where the sextupole is nested within the solenoids: the same geometry used in the VENUS magnet. In comparison to the other feasible magnet geometries, this magnet geometry results in the highest and the most complex interaction forces. However, it is a well-proven NbTi magnet design for the 3$^{rd}$ generation ECRIS.

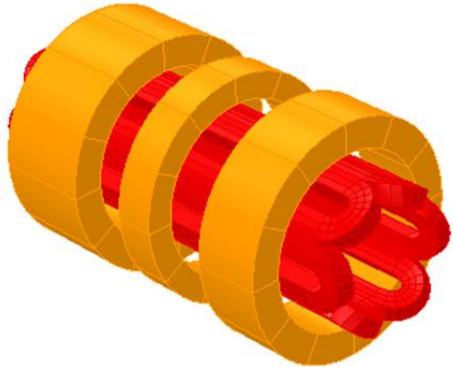

*Figure 2.1.4. The schematic layout of the Sextupole-In-Solenoids geometry used for NbTi-magnet-based VENUS and most of the 3$^{rd}$ generation ECRIS. The same magnet geometry is being used by IMP/Lanzhou/China for building a Nb$_3$Sn magnet for developing a 45 GHz ECRIS.*

As a Nb$_3$Sn magnet has never been developed and tested for an ECRIS, significant R&D effort is needed to address the daunting challenges related to this material, such as winding techniques, fixtures, magnet coil clamping and quench protection, etc. The ion source group of the Institute of Modern Physics (IMP) is aware of the challenging tasks in developing such a high field Nb$_3$Sn magnet for a 45 GHz ECRIS. Not long after the R&D started in 2014, they decided to prototype a half-sized Nb₃Sn magnet first to address some of the challenging fabrication issues. In the recently held 2021 International Conference on Ion Sources, they reported the completion of the prototyped half-sized Nb₃Sn magnet but achieved only 75% of the design solenoid field and 90% of the sextupole field. The inability to reach design fields with the smaller, simpler prototype magnet has demonstrated the difficulties in developing a high field Nb₃Sn magnet for ECRIS. However, IMP has decided to move forward to assemble the full-sized Nb₃Sn magnet and complete the cold tests in 2022. The remaining work on the Nb₃Sn magnet will be difficult and crucial as the magnet operation and long-term reliability need to be proven, even after the magnet capable of generating the designed fields.

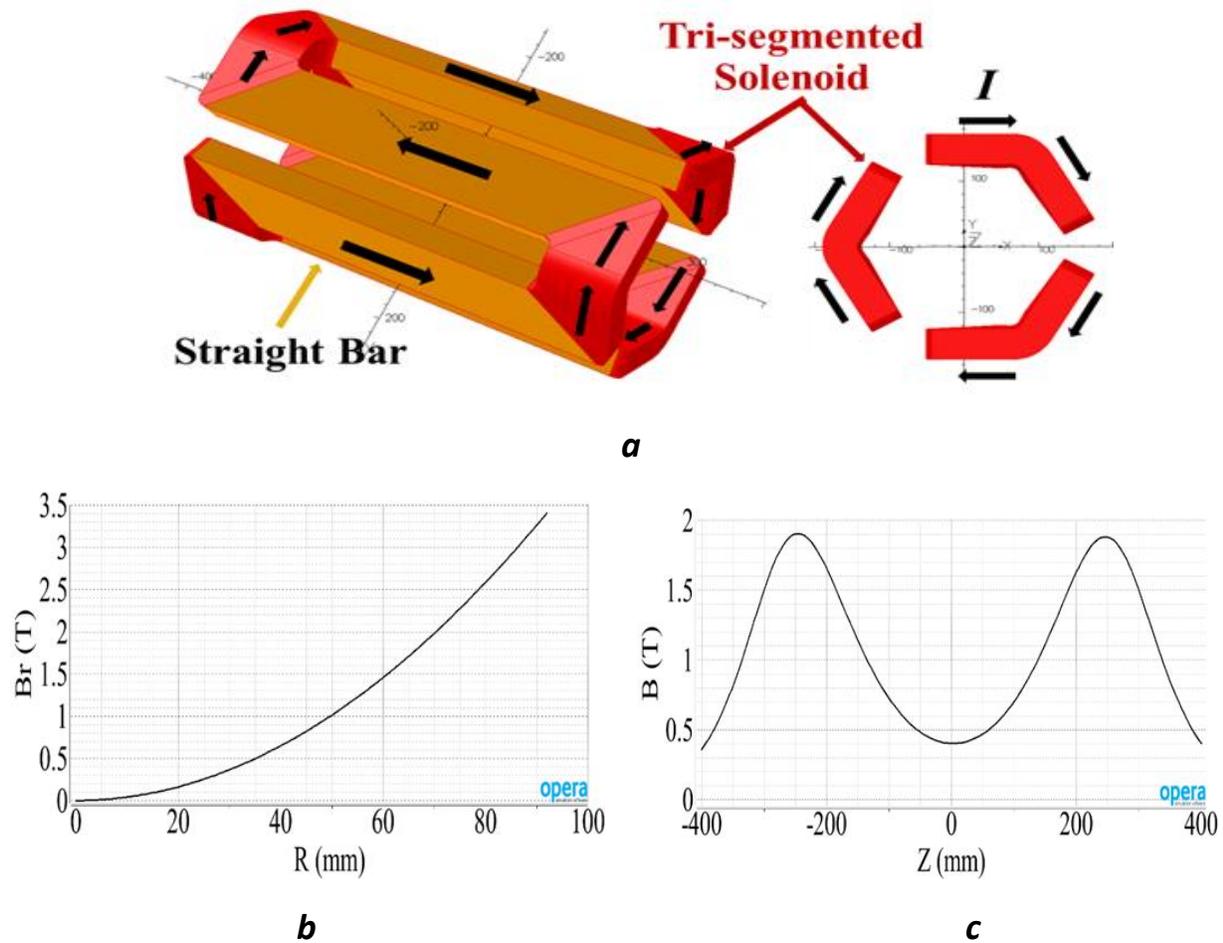

*Figure 2.1.5. **a**. Schematic layout of a closed-loop-coil, the MARS critical component, combining six rectangular straight bars (yellow color) and two three-segmented solenoids (red color). Shown in **b** is the radial field and in **c** the substantial axial field contribution calculated with OPERA-3D.*

*2.1.3.2. A novel MARS magnet geometry for future generations of ECRIS*

MARS, a Mixed Axial and Radial field System, [XIE2012], [XIE2015] is the alternative path with an optimized magnet geometry to obtain as high field as possible within the conductor constraints. The critical component of MARS is a closed-loop-coil with a hexagonally-shaped cross section, as schematically shown in Figure 2.1.5a. The long, straight sections of this geometry provide a sextupole field in a similar manner as the long sections of typical racetrack windings. The major difference between the MARS geometry and a conventional racetrack design comes in the connections between these long sections, in which the MARS closed-loop-coil end connections can be viewed as an optimization of coil field usage over the latter. In the MARS geometry, the currents in the sextupole ends rotate about the long axis in only one direction, whereas neighboring ends in the racetrack design have rotations that oppose one another. This single-direction rotation means that the closed-loop coil is providing a solenoidal field at each end (unlike the racetrack whose solenoid components cancel), reducing the needed strength of additional solenoid coils to reach required source fields. As can be seen in Figure 2.1.5b-c, the MARS closed-loop-coil winding alone produces a minimum-*B* field configuration that needs relatively little augmentation.

The shape of the plasma chamber is crucial to achieve the needed high radial fields in this new coil geometry. A hexagonal plasma chamber matching the shape of the closed-loop-coil would effectively increase available radial fields by about 30% as indicated in Figure 2.1.6, compare to that of a circular plasma chamber.

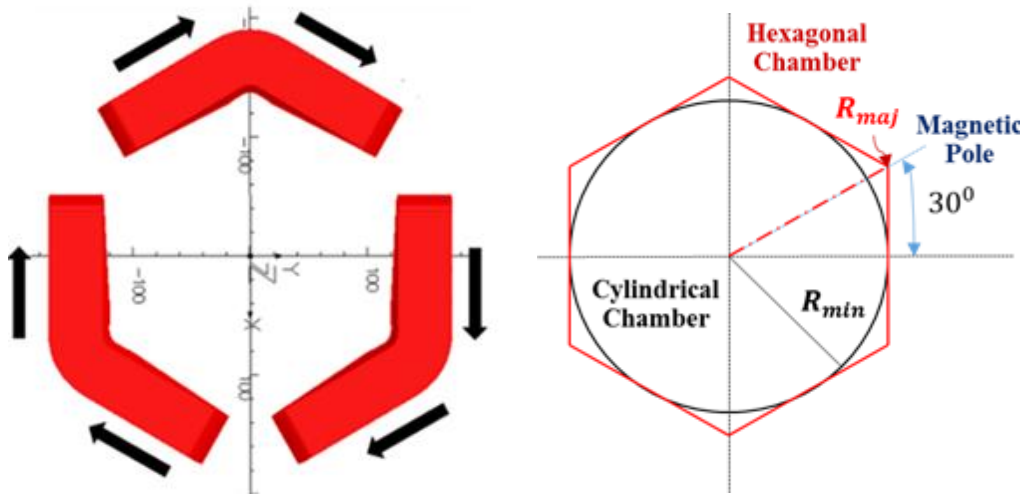

*Figure 2.1.6. A hexagonal plasma chamber matching the shape of the closed-loop-coil would yield higher radial field by about 30% in comparison to the typical cylindrical plasma chamber used in ECRIS, as $R_{maj} = R_{min}/\cos(30°)$ and the radial field $B_r$ is proportional to the square of the radial distance $R^2$.*

In summary, the MARS magnet geometry, in which the closed-loop-coil is a geometric optimization of the conventional racetrack sextupole coils, has a number of significant advantages over the existing ECRIS magnet schemes:

- **Simplified interaction force pattern and clamping.** By eliminating the racetrack coil ends creating opposing solenoid fields, the repulsive forces between the closed-loop-coil ends and the solenoids are converted into all attractive forces. The coil-end interaction forces are radially and axially outward result in a simpler coil clamping.
- **Azimuthally deformation-resistant.** After epoxy impregnation, a closed-loop-coil is structurally stronger at resisting azimuthal deformation due to the strong radial fields from the solenoids.
- **Higher radial field values.** Compared to a conventional sextupole, 30% higher radial fields can be achieved for the same number of ampere-turns.
- **Significant axial field contributions.** A MARS closed-loop-coil can contribute up to 30-40% of the axial peak fields needed for plasma containment. This results in substantially smaller solenoids to achieve the same peak axial fields.
- **A smaller, lighter magnet system.** Since the repulsions are eliminated, the solenoids can be placed just inside/outside or right next to the closed-loop-coil ends which leads to a shorter magnet (a length reduction of up to 30-40%).

Based on this novel magnet geometry and operating within the NbTi conductor constraints, as shown below in Figure 2.1.7, a NbTi MARS magnet could produce field maxima of ~ 5.7 T axially and ~ 3.2 T radially. Using only about 50% of the NbTi conductor necessary for VENUS, this magnet can produce about 40%-50% higher fields than that source and the minimum-*B* field strengths are sufficient for operating at 45 GHz [XIE2015]. As mentioned above, the closed-loop-coil is the sole technical hurdle of the MARS magnet geometry which results in the most challenging task in coil fabrications. Employing a set of winding fixtures and procedures designed for the task, a prototype of ~ 1/3 radial thickness of the closed-loop-coil using copper wires has been conducted and the field profiles verified.

After the successful copper closed-loop-coil construction and testing, a proposal titled "Development of a 4th generation NbTi MARS-magnet-based 45 GHz Electron Cyclotron Resonance Ion Source" was submitted to and reviewed by DOE Office of Nuclear Physics in 2021. This 45 GHz ECRIS is named MARS-D, and is a demonstration of the new MARS magnet geometry. If this proposal is funded, there will be an opportunity to pursue a novel NbTi MARS magnet for an ECRIS operation at 45 GHz. This would be another large step in ECRIS technology advancement and would be expected to have at least as great an impact on the nuclear physics community as VENUS has.

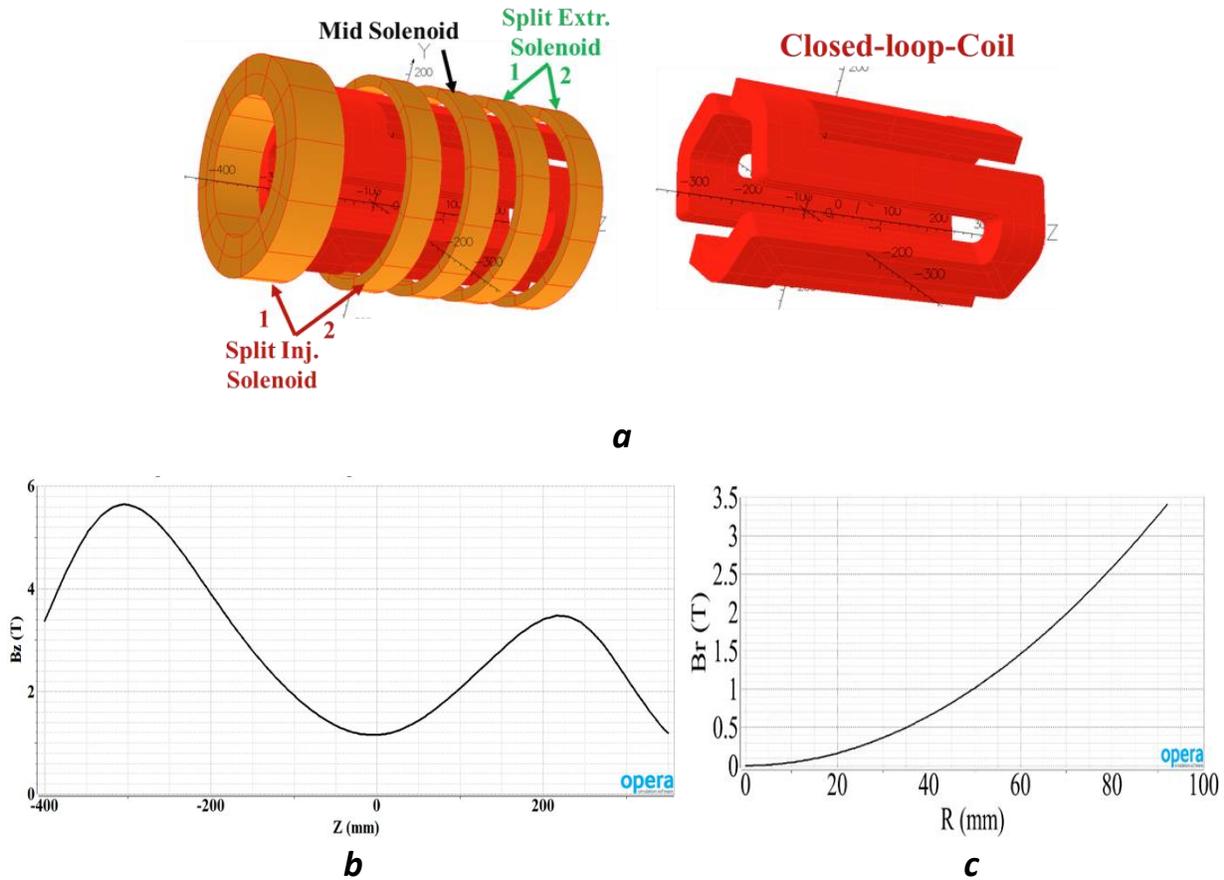

*Figure 2.1.7. **a.** Schematic layout of a NbTi MARS magnet for a next generation 45 GHz ECRIS. Shown in **b** is the axial field reaching 5.7 T and **c** is the radial field up to ~ 3.2 T at the chamber walls, calculated with OPERA-3D.*

Furthermore, the MARS magnet geometry can be applied to produce higher fields for a future generation of higher frequency ECRIS with higher-current conductors. Calculated fields for a MARS winding using high-temperature superconductor (HTS) REBCO conductors have shown fields of B ≥ 10 T on axis and Br ≥ 6 T at the plasma chamber walls are attainable within the conductor constraints. These fields would be high enough for ECRIS operation at ≥ 80 GHz. Additionally, the ~50% in conductor use for a MARS magnet as compared to a VENUS-type magnet geometry would lead to substantial conductor cost savings, especially for the future HTS magnets.

Table 2.1.2 and Figure 2.1.8 present the possible performance of the future generations of ECRISs. These extrapolations are made based on the performance of the past generations of ECRIS to yield some guidance for the conceptual design of future accelerators and possible future upgrades to the existing accelerators. These performance predictions are well justified for supporting future R/D on ECRIS.

*Table 2.1.2. A few typical cw beam intensities and enhancements from ECRIS*

| Ions | A/Q | Year Intensity ECRISs | Year Intensity ECRISs | Extrapolated Current I* (45 GHz & > 2024) |
|---|---|---|---|---|
| $O^{6+}$ | 2.67 | 1974<br>~ $1.5 \times 10^{13}$ ions/s<br>Supermafios | 2015-2019<br>~ $7 \times 10^{15}$ ions/s<br>LBL VENUS (18+28 GHz)<br>IMP SECRAL (18+28+45 GHz) | $\geq 1 \times 10^{16}$ ions/s |
| $Xe^{30+}$ | 4.30 | 1997-1998<br>~ $3 \times 10^{12}$ ions/s<br>RIKEN 18 GHz<br>LBL AECR-U (10+14 GHz) | 2015-2019<br>~$7 \times 10^{13}$ ions/s<br>LBL VENUS (18+28 GHz)<br>IMP SECRAL (18+28+45 GHz) | $\geq 2 \times 10^{14}$ ions/s |
| $Bi^{35+}$ | 5.97 | 1997<br>~ $2.6 \times 10^{11}$ ions/s<br>LBL AECR-U (10+14 GHz) | 2019<br>~ $1.2 \times 10^{14}$ ions/s<br>IMP SECRAL (18+28+45 GHz) | $\geq 2.5 \times 10^{14}$ ions/s |
| $U^{34+}$ | 7.00 | 1997<br>~ $3.5 \times 10^{12}$ ions/s<br>LBL AECR-U (10+14 GHz) | 2011<br>~ $7.0 \times 10^{13}$ ions/s<br>LBL VENUS (18+28 GHz) | $\geq 2 \times 10^{14}$ ions/s |

*\* The extrapolated performances are for the under-development next generation 45 GHz ECRIS with heating power of ~ 20 kW.*

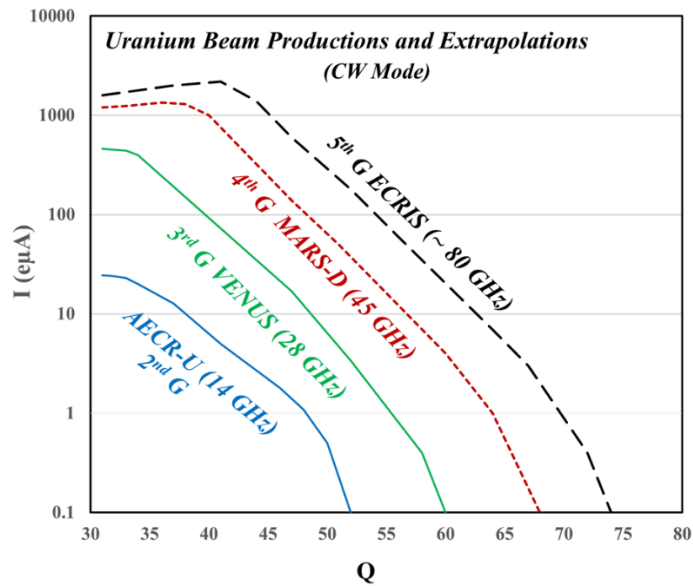

*Figure 2.1.8. Extrapolated uranium beam performance of a 4$^{th}$ generation 45 GHz ECRIS (MARS-D) and a likely 5$^{th}$ generation 80 GHz ECRIS, scaled from the performance of the 3$^{rd}$ generation 28 GHz ECRIS VENUS and the 2$^{nd}$ generation 14 GHz AECR-U.*

### 2.2 Laser Ion Source (Lead Contributor: **Okamura**)

*2.2.1. Introduction*

A laser ion source (LIS) is a simple, powerful, and high brightness pulsed ion source. A LIS has many advantages and is potentially being adopted by many accelerator facilities. Due to its very simple configuration, a LIS matches not only to large-scale facilities but also to small accelerators. A LIS has several advantages listed below.

1. The ionizing energy is given only by laser light. We do not need to have a power source at a high voltage terminal. A LIS can be easily mounted at the terminal, including electrostatic accelerators.

2. Ionization occurs in the nanosecond range and the plasma cannot expand much within the ionization period. This means that we do not need to provide any plasma confinement forces.

3. The beam current can easily reach more than 100 mA with a tabletop laser system. It is probably the most powerful heavy ion source at a reasonable budget.

4. A very high charge state can be effortlessly achieved from light- to medium-mass species.

5. The ionizing plasma volume is very small and far away from the extraction electrode. Thus, we can obtain a very uniform beam from well-cooled expanded plasma. This feature enables us to extract a minimum emittance beam with great uniformity.

6. Due to the speed of the plasma moving, we can extract a more intense beam than that predicted by the static law of three halves.

7. The direct plasma injection scheme (DPIS) can be applied.

Of course, there are some drawbacks.

1. The duration of the beam pulse is from one to a hundred microseconds only. Only pulsed-ion beams can be delivered.

2. In the case of high charge-state production mode, the ion beam has momentum spreads (but this can be compensated).

3. To achieve high charge state ions from very heavy materials, a powerful laser system is demanded.

Since the first idea of an ion source using laser ablation plasma arose, almost fifty years have already passed. However, LIS has not yet been widely spread out. We hope that this White Paper will encourage more scientists and engineers to become familiar with laser ion sources.

In this article, we only discuss laser ablation plasma ion sources. A selective LIS, which uses a resonant ionization process, and a Target Normal Sheath Acceleration (TNSA) LIS are excluded.

## *2.2.2. The basic principle of LIS*

A typical laser ion source consists of a laser system, a target, and an extraction electrode as shown in Fig. 2.2.1. Here we assume that the laser pulse duration is in the range of 5 to 10 ns with the highly focused condition. The spot size is typically 10 to 100 μm to obtain highly charged ions. In the very early part of the laser irradiation period, the laser energy is not absorbed efficiently. Once an initial plasma is formed, however, the laser energy starts being converted to the electron temperature by the classical absorption process, and then the plasma is rapidly heated. As a result of collisions between ions and electrons, stepwise ionization occurs. Simultaneously, the plasma starts expanding. At the end of the laser irradiation period, the front end of the plasma reaches a few mm from the target surface, which becomes larger than the laser spot size. After the laser irradiation period, the plasma keeps expanding three-dimensionally and becomes colder and colder. The entire expanding plasma moves away from the target surface. In the case of high charge state ion production, the velocity of the plasma front reaches around 100 eV/u. When one prefers to have lower charge-state ions, the laser spot size can be increased up to a few mm size by changing the laser focusing condition and the plasma temperature is lowered. The heating stage plasma's shape may be very thin because the plasma expansion becomes slower and the plasma front proceeds less than 1 mm. If the plasma is optimized to provide singly charged ions, the velocity of the plasma front reaches approximately 1 eV/u.

Now, we pay attention to the trajectory of each ion. At the plasma heating stage, the size of the formed plasma can be assumed as a pinpoint, because its size is negligible compared to the drift length shown in the figure. Therefore, in the plasma expansion stage, all the ions start from the identical point in the space at the surface of the laser target. Then, each ion moves straight with a constant velocity while drifting, and the expansion occurs in a three-dimensional space. The difference in arrival times of the fastest and slowest ions at the extraction point defines the ion beam pulse width. Although the laser irradiation is very short (~10 ns), the pulse width of the extracted ion beams can be extended to the microsecond scale.

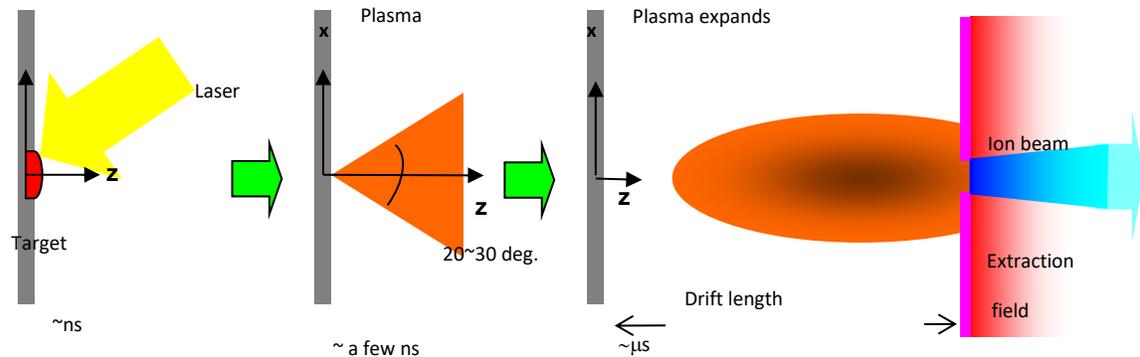

*Figure 2.2.1. Principle of a laser ion source.*

### 2.2.3. General description

*2.2.3.1. Laser system*

Historically, $CO_2$ lasers have been used in many institutes. The advantage of the $CO_2$ laser is its high energy output and cooling capability. It can be operated even in cw mode and is widely used in the industrial machining field. An example of the most powerful $CO_2$ LIS was demonstrated in 2003 by the CERN and ITEP group. They developed a gigantic 100 J laser system [SHAR2005] that could deliver lead ions with charge states of 19 + to 32+ and charge states 26+ and 27+ showed the highest yield.

A typical wavelength of a $CO_2$ laser is about 10 µm, which is in the infrared spectral region. Therefore, a vacuum window made from zinc selenide or salt crystal is used, which is transparent for the wavelength. A $CO_2$ laser has a medium gas mixture and requires a discharge to obtain a population inversion. Due to the discharge process, special attention is required to obtain good stability. The pulse length is typically more than a few tens of ns. Within this relatively long laser pulse period, the laser continuously transfers energy to the plasma near the target surface, hence the plasma does not get heated evenly. Therefore, the momentum distribution of the ions in the plasma does not represent a shifted Maxwell-Boltzmann distribution. A $CO_2$ laser is one of the candidates for driver lasers.

We have used Q-switched Nd-YAG lasers for ion source applications for more than ten years. Many reliable models in a reasonable cost range are available on the market. The fundamental wavelength and typical pulse duration are 1064 nm and 6~10 ns respectively. The laser energy can be easily controlled by changing the interval between the flash lamp trigger and Q-switch timing or the flash lamp excitation voltage.

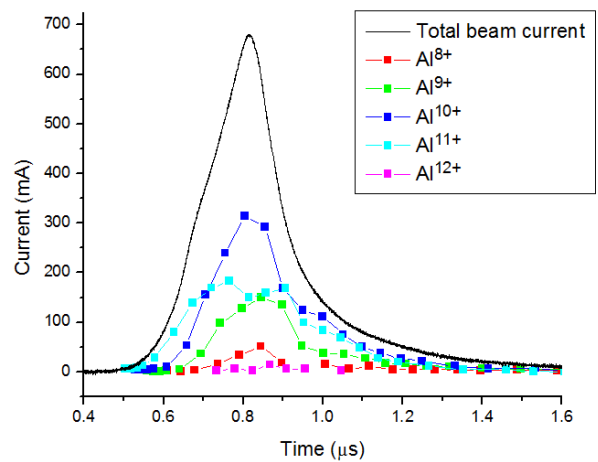

*Figure 2.2.2. Current profile of aluminum charge states produced by a 6 ns, 1064 nm, 840 mJ Nd-YAG laser.*

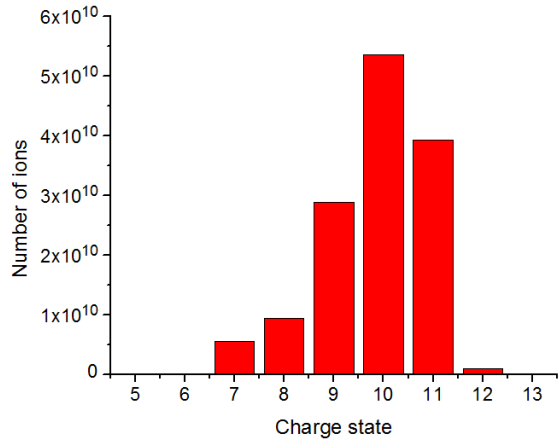

*Figure 2.2.3. Charge state distribution of aluminum produced by a 6 ns, 1064 nm, 840 mJ Nd-YAG laser.*

To minimize undesired target damage, a contrast value of the Q-switch is important so that a laser leakage before opening the Q-switch may heat the target before starting the main laser pulse. The major optic parts, including lenses, mirrors, and Faraday isolators, can be obtained at reasonable costs. Figures 2.2.2 and 2.2.3 show the typical ion beam profile and charge state distribution of aluminum plasma using 6 ns, 1064 nm, 840 mJ Nd-YAG laser. The beam corrector with a 5-mm aperture was placed 300 mm away from the target surface.

We have tested shorter pulse length lasers including a sub-nanosecond laser system, which has a stimulated Brillouin scattering (SBS) cell. Some reports indicate that the laser power density around $10^{14}$ W/cm$^2$ may be able to produce very high charge state ions [LASK2005, LASK2004] and the sub-nanosecond laser was supposed to provide a close value to the singular working point. Unfortunately, so far, we have not found significant advantages compared to a typical Nd-YAG laser. We could observe a small amount of higher charge state ions, but the total quantity of the plasma reduces one order less than that of the plasma created by typical nanosecond lasers. We suppose that the laser pulse length maybe not be enough to achieve temperature equilibrium conditions in the plasma heating stage.

*2.2.3.2. Target*

A LIS can provide a very wide range of species except for helium. All solid materials can be ionized by laser irradiation. In Brookhaven National Laboratory (BNL), we regularly provide Li, B, C, O, Al, Ca, Si, Ti, Fe, Zr, Nb, and Au to the user facilities from a LIS. Proton and oxygen beams can be easily obtained from Zr-hydride and alumina targets. We have also demonstrated how to obtain good beams from frozen Ne and Ar targets using a cryocooler head.

Due to laser ablation, the material at the target surface is consumed. In the case of high-charge state production mode, we need to focus the laser beam at the surface. The material that contributes to creating high-temperature plasma is at only the surface layer of the target, which is typically less than 500 nm in thickness. However, heat from the plasma induces subsequent ionization and evaporation of the material from the deeper layer, and a crater is formed which may reach down to 200 μm of depth. Once a crater is formed, we cannot apply a second laser shot on the crater since the focal spot of the laser is in the space in the crater and the effective laser power density is decreased. It also influences beam stability. To avoid these negative effects, the target needs to be scanned to provide a fresh target surface for every laser shot.

For the low charge-state production mode, laser power density on the target surface should be controlled to have desired charge state in the plasma. For example, if we would like to have a 1+ charge state beam, the laser power density should be adjusted between $2 \times 10^8$ and $10^9$ W/cm$^2$ for efficient ion production. The laser spot size on the target surface would be several mm in diameter when we use several hundred mJ of laser energy. In that case, the damage on the target surface caused by single laser irradiation is minor, and we can apply multiple shots on the same spot up to several hundred times. After many irradiations, the surface becomes damaged by ablation. The remaining melted surface layer after laser irradiation is rapidly resolidified. During the very short time that the target material is liquefied, surface tension causes blisters formation. To minimize the effect of blisters, the target can be slowly scanned for long-term maintenance-free operation.

### *2.2.3.3. Plasma drift length*

As mentioned above, an ablation plasma plume expands in three dimensions in space. When the head of the plasma plume reaches the extraction voltage gap, ion beam formation occurs. Ion beam generation continues until the end of the plasma plume reaches the extraction electrode. If the distance from the target to the extraction point is extended, the plasma expands more. Therefore, more distance makes for a longer ion pulse and a thinner ion density. We call the distance "plasma drift length". Plasma drift length is an important parameter to characterize the ion beam properties. The ion pulse width is proportional to the length and the peak's current amplitude is inversely proportional to the cube of this value. Figure 2.2.4 shows the relationships of peak currents and pulse widths as a function of distance for two different target materials and charge states.

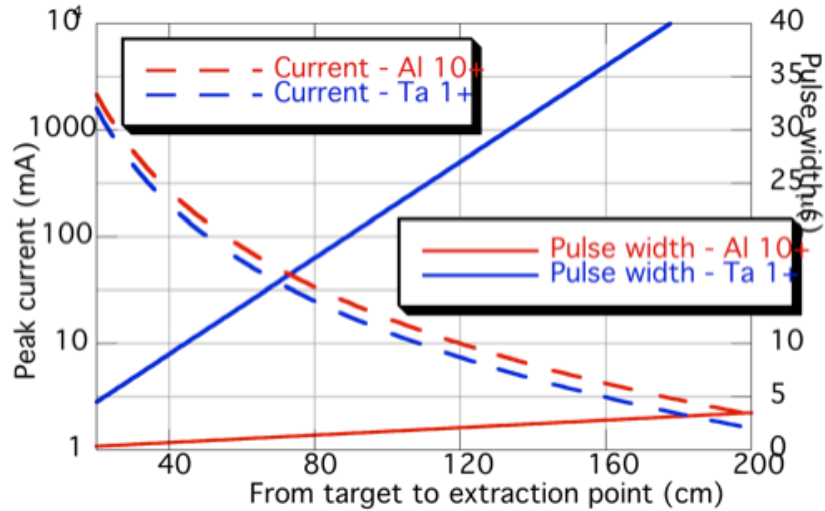

*Figure 2.2.4. Ion beam peak currents and ion beam pulse width vs. plasma drift length.*

### 2.2.3.4. Ion beam extraction

The expanded plasma moves to the beam extractor, and the strong electric force pulls the ions out of the plasma at the extractor. It is well known that the maximum beam current is limited by the space charge. For a static plasma, the space-charge limit is expressed by the famous three halves law as,

$$J = \frac{4}{9} \varepsilon_o \sqrt{\frac{2q}{m}} \frac{V_a^{3/2}}{d^2}$$

In the formula, $J$, $\varepsilon_a$, $q$, $m$, $V_a$, and $d$ are the limiting current density, permittivity, the charge of the particle, the mass of the particle, the applied voltage between the extraction electrodes, and the distance of the electrodes, respectively. This formula is derived from the Poisson equation with zero initial velocity of the charged particles. In the case of an LIS, the ions already have velocity toward the extraction region and this makes a slight difference in the formula [JAFF1944, BENI2009, LIU1995],

$$J = \frac{4}{9} \varepsilon_o \sqrt{\frac{2q}{m}} \frac{\left(\sqrt{V_o} + \sqrt{V_o + V_a}\right)^3}{d^2}$$

where $V_0$ is the voltage corresponding to the initial velocity of an ion in the sheath at the starting electrode. In the case of high charge state ion production, the $V_0$ goes up to a significant value and the maximum beam current can be increased several times.

For rapid cycling beam extraction, the evacuation system at the extractor needs to be carefully designed to achieve and maintain good vacuum conditions. In addition to the main plasma pulse, a neutral vapor reaches the extraction area that may cause electron recombination of ions and also may trigger discharges. A sample design of the LIS extraction system is illustrated in Fig. 2.2.5. This is being used to provide low charge state ion beams in BNL and the measured typical emittance is 0.043 mm mrad (RMS normalized, $Au^{1+}$ beam).

Here, we would like to emphasize that LIS does not have a background magnetic field. This condition helps to minimize beam emittance from a LIS.

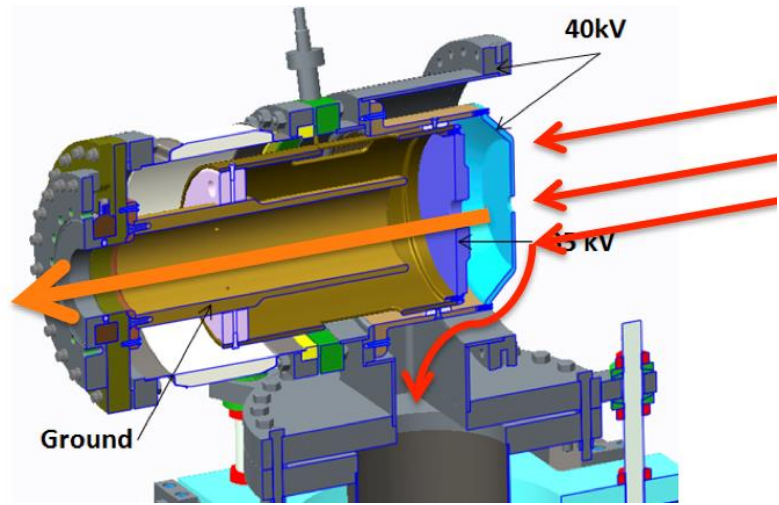

*Figure 2.2.5. Ion beam extractor used in the low charge state LIS in BNL. The system consists of three electrodes. The first gap determines the extraction voltage and the second gap is used to achieve the required platform voltage. This source is not equipped with a back-stream electron suppression electrode.*

*2.2.3.5. Direct plasma Injection scheme*

We have originally developed a unique technique called the Direct Plasma Injection Scheme (DPIS) since 2001 [OKAM2002]. A LIS can provide an exceptionally intense beam, nevertheless, the beam loss at the beam transport line between the LIS and first stage accelerator restricts usable beam currents. This is due to strong space charge force with high current and low-velocity transport conditions. DPIS can overcome this issue. The DPIS consists of a LIS and a Radio-Frequency Quadrupole (RFQ) linear accelerator. As we know, the ablation plasma has an initial velocity normal to the laser target surface. This means that we do not need extracted beams to transport to an RFQ, which is commonly used as a first-stage accelerator. Because plasma travels by itself, the ions contained by the plasma can be transported as a neutral state. Then, the plasma is directly injected into the RFQ cavity and the ion beam is extracted.

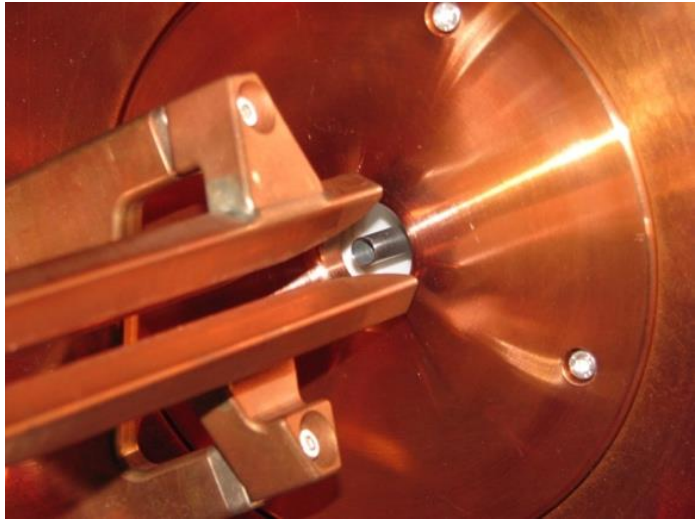

*Figure 2.2.6. Plasma injection point.*

Figure 2.2.6 shows the inside of the RFQ cavity. The stainless steel nozzle is located at the beam axis of the RFQ. The nozzle is isolated, and an injection voltage is applied to it. The beam emission surface is slightly inside of the nozzle. The nozzle is mechanically and electrically connected to the high voltage box, which is surrounded by the grounded vacuum vessel as shown in Fig. 2.2.7. The laser light is guided and focused on the target material in the high voltage box and the induced plasma expands in the space enclosed by the same potential metal wall up to the end of the nozzle. The applied high voltage is not exposed to the outside of the ion source. Therefore, neither a safety cage nor a platform is required.

As mentioned, the DPIS eliminates the space-charge effect at the low energy beam transport line and the high brightness beam can be effortlessly transferred to an RFQ which has strong transverse focusing force. DPIS has another advantage to handle high current beams. Figure 2.2.8 shows equipotential lines in an RFQ with DPIS. The intervals of the equipotential lines are very dense at the edge of the nozzle. Therefore, the electric field strength at the beam extraction region is much enhanced compared to orthodox parallel plate shape electrodes. Using DPIS, peak current after the RFQ can be achieved more than 100 mA easily.

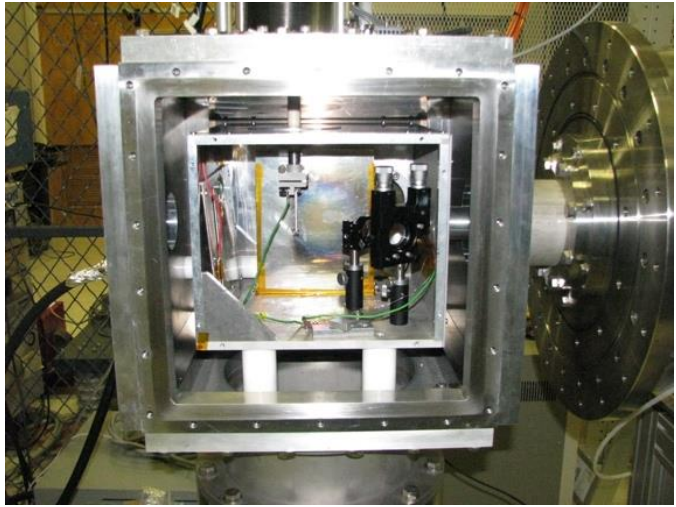

*Figure 2.2.7. Laser illumination box.*

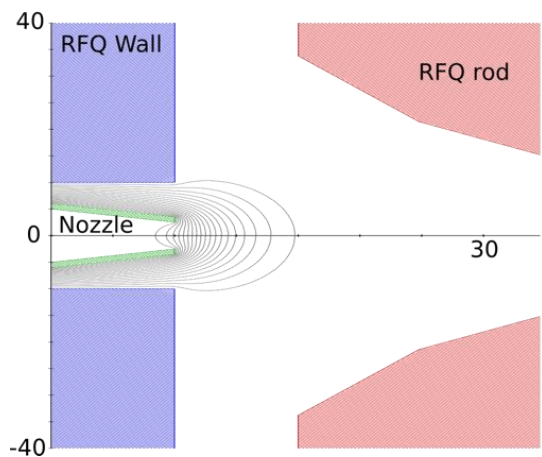

*Figure 2.2.8. Cross-sectional view of typical DPIS set-up with equipotential field lines.*

### 2.2.3.6. Beam current manipulation

A LIS has a very simple structure. In other words, it does not have many knobs to adjust the ion beam property. The charge-state distribution can be adjusted by the focal condition of the laser light or laser energy. For instance, defocusing the laser or reducing the laser energy can lower the most abundant charge state. There had been not many adjustment procedures in an LIS when the system was in operation.

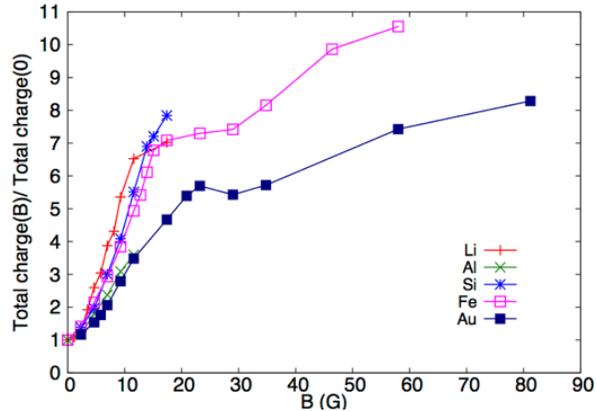

*Figure 2.2.9. Total charge enhancement.*

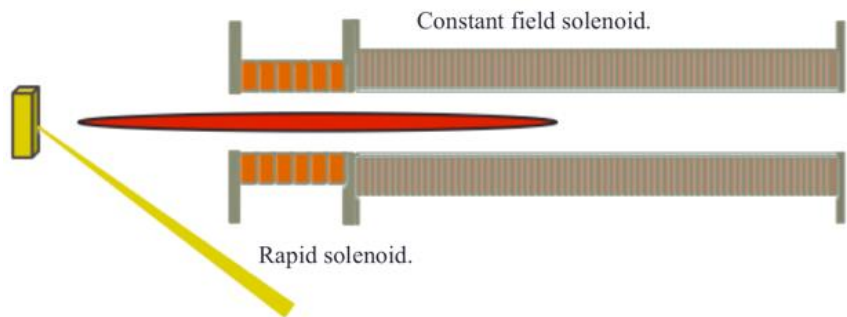

*Figure 2.2.10. Schematic view of solenoids.*

In 2009, we established a beam current enhancement technique by applying a weak solenoid field to the plasma drift section [OKAM2010]. When the plasma is in an expanding stage, it spreads three-dimensionally. However, by overlaying an axial magnetic field on the expanding space, the transverse expansion of the plasma can be restricted.

Figure 2.2.9 shows the effect of a static solenoid field, which is being used in the low charge state LIS of BNL. The solenoid was wound around the plasma drift pipe, which has 76 mm of inner diameter and 3.0 m in length. This is a very convenient knob to adjust the current of the entire beam pulse. In the LIS, another short solenoid was recently installed to control the beam current profile, as shown in Fig. 2.2.10. This short solenoid is ramped up to 60 Gauss (6 mT) in 10 µs. The inner diameter and length of the short rapid solenoid are 75 mm and 56 mm, respectively. This solenoid can enhance only within a certain time slice of an ion beam pulse. Therefore, this is another convenient knob to tailor a beam current profile [SEKI2015, OKAM2016].

Figure 2.2.11 shows a set-up of a long solenoid for the DPIS acceleration test using a 1 J, 6 ns, 1064 nm Nd-YAG laser. We tested carbon beam production. The inner diameter of the solenoid is 76 mm, and other dimensions are indicated in the figure. At 900 Gauss (90 mT) of solenoid field, $C^{4+}$ and $C^{6+}$ were accelerated to 100 keV/u. A single laser shot provided 36 mA at peak, 2.1 μs of pulse width $1.2 \times 10^{11}$ particles $C^{4+}$ [KANE2014] and 33 mA at peak, 1.6 μs of pulse width $5.2 \times 10^{10}$ particles $C^{6+}$ after the RFQ. The total accelerated beam current could be easily adjusted by changing the solenoid field strength. The charge-state switching was done by finely adjusting the target position. The solenoid is a very effective tool for controlling the plasma expansion in the LIS.

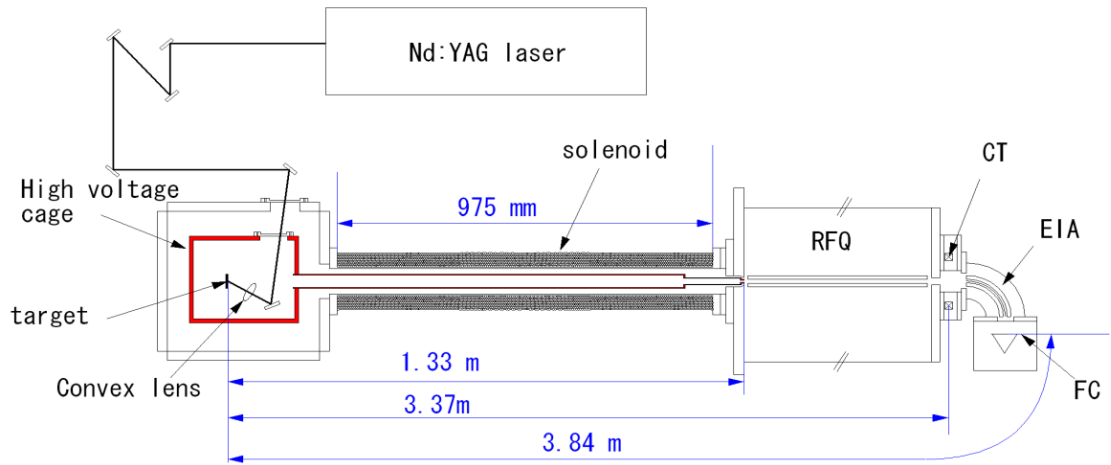

*Figure 2.2.11. DPIS setup for carbon acceleration with the solenoid.*

### 2.3. Charge breeders (Lead Contributor: **Lapierre**)

Charge breeders convert beams of ions of low charge state (mostly 1+ or 2+) into highly charge ion beams (HCIB). They were introduced more than 20 years ago in accelerator systems in the context of rare (radioactive) ion beam (RIB) acceleration after production. As the energy of ions accelerated through an accelerator scales with their charge, charge breeders are employed at accelerator facilities to extend the energy range. RIB are of interest in nuclear physics to expand our knowledge of nuclear structure as well as in nuclear astrophysics to understand the nucleosynthesis of heavy elements. At RIB facilities, the rare isotopes produced by, for instance, fast projectile fragmentation or the ISOL (Isotope Separation On-Line) technique are charge bred for (post-) acceleration to reach energies that differ from the energy of the rare isotopes after production (see review in [LAPI2019]). Rare isotopes are produced in minute quantities and can have short decay times of less than 100 ms. Efficient and fast production of high-quality HCIB, free of contaminants, is critical at any post-accelerator facility. Electron Cyclotron Resonance Ion Sources (ECRIS) and Electron-Beam Ion Sources or Traps (EBIST) are employed at post-accelerator facilities (see Fig. 2.3.1) as charge breeders for their highest stripping efficiencies and beam qualities. In an ECRIS breeder, injected ions are confined in an electron-ion plasma and ionized by plasma electrons. They have a large charge capacity for efficient capture of intense beams and typically operate in a continuous mode of beam injection/extraction suitable for efficient detection of events after nuclear reactions. However, they have a high pervasive contaminating background that can overwhelm RIB. EBIST breeders uses a magnetically compressed (monoenergetic) electron beam to breed injected ion pulses from an upstream cooler-buncher ion trap. They have a significantly lower background than ECRIS breeders. However, their charge capacity is limited to beams of low intensity and the high instantaneous rate of the extracted short ion pulses can impede efficient ion detection. EBIST breeders are often considered the device of choice due to their high breeding efficiencies and low background, but the small charge capacity of EBIST breeder systems can be a bottleneck at future RIB facilities expected to increase the production rates of radioactive ions by several others of magnitude [FRIB2022]. Future RIB facilities may necessitate the implementation of both systems, having complementary operating capabilities. The strengths, weaknesses, capabilities, and paths forward for improving the performance of ECRIS and EBIST breeders is discussed in the context of future (high intensity) accelerator facilities.

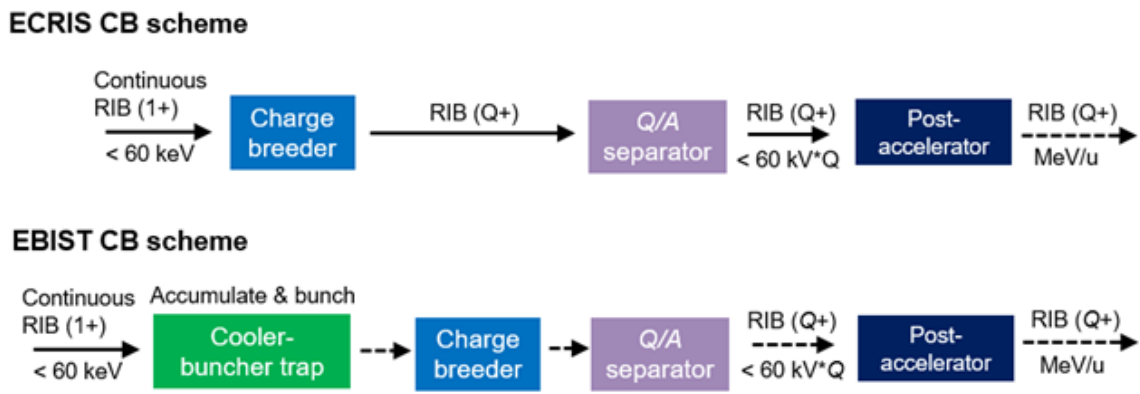

*Fig. 2.3.1. Typical post-acceleration scheme used with ECRIS and EBIST breeders employed at post-accelerator facilities. Ion beams (mostly 1+ or 2+) are injected into ECRIS breeders and extracted in a continuous mode. Ion pulses are extracted from EBIST breeders after pulsed injection from an upstream cooler-buncher ion trap accumulating continuous beams. All systems incorporate downstream of the breeder, a separator for charge-over-mass (Q/A) selection before injection into the post-accelerator (e.g., linear accelerator (LINAC) or Cyclotron) for subsequent acceleration up to several to tens of MeV/u.*

### *2.3.1. EBIST breeders*

*2.3.1.1. Design and operation*

The design of EBIST breeders is based on the same basic principle as those employed as ion sources or traps [CURR1995]. This is shown in Fig. 2.3.2. They produce and confine HCI with an electron beam compressed by a magnetic field of high flux density. They are composed of an electron gun, a superconducting magnet solenoid (or Helmholtz coils), coaxial cylindrical electrodes placed in the magnet bore, and an electron collector. The electron beam is injected (on-axis) into the magnetic field and the central axis of an electrode structure where ions are axially trapped by two electrostatic potential barriers and radially confined by the electron-beam space-charge potential. The trapped ions are ionized by the electron beam to high charge states by successive single ionization. After crossing the trapping region, the electron beam is stopped within the collector. EBIST breeders are designed to maximize the trapped-ion charge capacity, injected-beam acceptance, and electron-current density with a high electron-beam current of up to 3 A and a long trap structure ($\lesssim$ 0.75 m). The charge capacity is typically less than $10^{11}$ elementary charges (1 charge = 1.602 $10^{-19}$ C). A high acceptance is needed to efficiently capture ion pulses having large longitudinal and transverse emittances (that can often result from the injection of bunched ion pulses). High magnetic compression of a high-current electron beam is technically challenging and can also reduce the transverse acceptance, proportional to the electron-beam size. Hence, EBIST breeders reach high current densities ($\lesssim$ 2000 A/cm$^2$) by only moderate compression of the high-current electron beam. The electron density is sufficient to breed *A/Q* ~ 2 (*A* is the atomic mass number and *Q* is the ion charge) for light ions and *A/Q* ~ 3 - 7 of heavy ions within less than hundreds of ms. Table 2.4.1 lists typical operational parameters/characteristics of EBIST breeders in operation and under construction (REX ISOLDE [WENA1999], ReA [LAPI2018], CARIBU/ATLAS [DICK2018], CANREB [BLESS2015], and RAON [SON2017]).

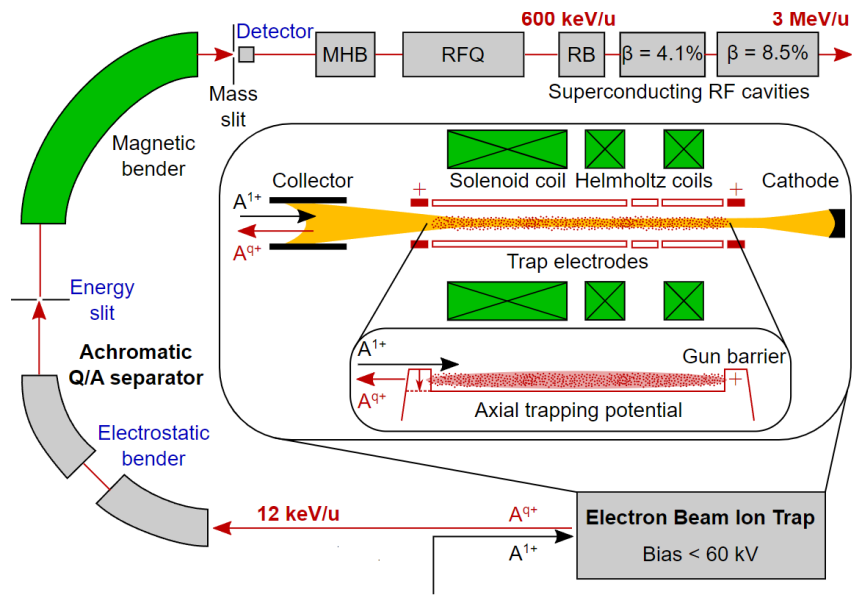

*Figure 2.3.2. Drawing of the ReA EBIST breeder and first section of the ReA post-accelerator (RFQ: Radio-Frequency Quadrupole accelerator, MHB: Multi-Harmonic Buncher, RB: Rebuncher). The high-current electron beam from the EBIST electron gun is magnetically compressed in the ion trap's center to a high*

*current density, needed for production of HCI by electron-impact ionization. Ion pulses are injected from an upstream cooler-buncher ion trap and extracted to charge-over-mass (Q/A) separator [BAUM2015].*

Pulsed injection is the preferred method for injection of beams into EBIST breeders. Ions of tens of keV in energy from a feeding ion source are injected as continuous beam into a cooler-buncher ion trap (gas-filled linear Paul or Penning trap). The ions are decelerated, accumulated, bunched, and cooled with a buffer gas (e.g., helium) to reduce the emittance of the ejected pulsed beam. The efficiency of bunchers in operation ranges between 20% and 100%, depending on the mass of the isotopes. Each ejected pulse of typically less than 5 µs in width is then transported at tens of keV to the EBIST. The ion pulse is injected through the collector (see Fig. 2.3.2.), decelerated to hundreds of eV, enters the trapping region after dynamically lowering a barrier potential below the beam energy, and then captured by raising this potential when it reaches the trap center. This pulse scheme allows for high capture efficiencies independently of the electron-beam current density. After breeding, a HCI pulse is then released by either lowering the barrier potential or raising the central trap potential and extracted by passing through the collector. EBIST breeders are pulsed machines. As such, the capture (breeding) time of the injected ion pulses can be varied to modify the charge-state distribution of the trapped ions and optimize production of specific charge states needed for post-acceleration. The combination of the breeding and extraction times defines the pulsed injection and extraction repetition frequency, normally less than 100 Hz. The efficiency of EBIST breeders operating in the pulsed mode can reach in single charge states of up 35% (excl. the buncher efficiency).

*Table 2.3.1. Typical operational parameters of EBIST breeders.*

| A/Q range | $\lesssim 7$ |
|---|---|
| Typical mass range | $\lesssim 200$ u |
| Breeding efficiency (excl. buncher eff.) | $\lesssim 35\%$ |
| Extract. ion-beam emittance (1x RMS) | $\lesssim 0.04$ mm mrad |
| ΔE (energy spread) | $\lesssim 300$ eV |
| Electron-beam current | 0.2 – 3 A |
| Electron-beam energy | 5 – 30 keV |
| Nominal magnetic field | $\lesssim 6$ T |
| Electron-beam radius | $\lesssim 600$ µm |
| Electron-beam current density | $\lesssim 2000$ A/cm$^2$ |
| Capacity (units of elementary charge) | $\lesssim 10^{11}$ |
| Length of the trapping region | $\lesssim 0.75$ m |

*2.3.1.2. Limitations and areas of development*

<u>High electron current and current density for high-intensity beams</u>

Present RIB facilities have typical production rates that do not exceed ~10 million particles per second (pps), which can translate into beam intensities available for post-acceleration of less than a million pps. Future RIB facilities, such as the Facility of Rare-Isotope Beams (FRIB) [FRIB2022], are expected to increase such rates by several orders of magnitude. For certain isotopes near the valley of stability, rates up to $10^{12}$ pps can be expected at full operating power, resulting up to ~$10^{11}$ pps available for charge breeding. Moreover, RIB of lower rates can be contaminated with isobars and other beam constituents of high intensities that cannot be mass separated before charge breeding, increasing the overall beam intensity. All EBIST breeders currently in operation achieve stable operation with an electron-beam current of less than ~1 A and a typical current density of ~500 A/cm$^2$. This current is sufficient to confine a maximum of ~$6\times10^{10}$ elementary charges, which corresponds to $6\times10^9$ particles for an ion charge state of $Q$=10+. The density of the electron beam defines the charge-breeding time and hence the operating (injection-extraction) repetition rate from the cooler-buncher to the post-accelerator. For short breeding times allowing efficient operation at a maximum rate of 100 Hz and assuming that the electron beam can be fully compensated by the charge of the trapped ions, present EBIST breeders are limited to breeding a maximum of $6\times10^{12}$ charges per second, ~$6\times10^{11}$ pps for $Q$=10+. Operation at such high repetition rates can be achieved with the current density of existing systems to reach high charge states of light ions. However, the maximum number of extracted ions is significantly reduced for heavy elements that have higher charge states and necessitate longer breeding times. For pulsed injection into EBIST breeders, the cooler-buncher is nevertheless the bottleneck. Cooler-bunchers built to handle high beam rates can efficiently operate with a few $10^8$ ions per pulse. At this high rate, the emittance of the pulsed beams can considerably increase, which can affect negatively the injection and charge-breeding efficiency. The charge capacity of such bunchers is proportional to the length of the trapping region. A 10-fold improvement in their capacity could possibly be achieved by lengthening the trapping region from several centimeters to nearly a meter. However, doing so would only allow for handling rates of up to ~$10^{11}$ pps of light ions with operation at 100 Hz.

To circumvent this problem, direct injection of (quasi-)continuous beams into EBIST breeders eliminates the use of an upstream cooler-buncher. This, in turn, can help reduce the injected beam emittance that can result from ion bunching and hence ease the need for a high electron current required for high acceptance. However, for continuous injection, the ions are injected with a kinetic energy higher than a trapping barrier potential. The ion capture efficiency relies on the probability that the ions be ionized to higher charge states (1+ to 2+ or higher) before a round trip in the trap. This requires a high-current density. For a half-meter long trapping region, an estimated current density of more than ~1000 A/cm$^2$ is needed to exceed a capture efficiency of ~70%. This density can typically be achieved with low electron currents of less than ~250 mA, but has not been clearly demonstrated for currents higher than ~ 0.5 – 1 A. Such a high current is needed to keep the acceptance larger than the injected beam emittance to guarantee an appreciable overlap with the electron beam. With this current density and higher, reaching the EBIST charge-capacity limit is then less of a concern because charge breeding can be performed at high repetition rates for light as well as heavy elements.

The development of high current, high current-density electron guns for EBIST breeders is needed for efficient charge breeding of high-intensity beams. This would enable two operating modes: 1) pulsed injection at high repetition rates (short breeding times) to remain within the buncher's charge-capacity limit and 2) continuous injection with longer breeding times for beam rates exceeding the buncher's capacity. A long-term target goal of 4 A and 5500 A/cm$^2$ can be suggested. As examples, based on charge-breeding simulations, such a high current and current density would allow for efficient charge breeding within 10 ms (100 Hz) of 10$^{12}$ pps of He-like ions of mid-$Z$ elements such as argon ($Z$ is the atomic number) and Ne-like ions of heavy elements such as krypton. Promising work in this direction has been performed over the years [PIKI2015, MERT2017, PAHL2022], but development is still needed.

The production of increasingly high charge states in EBIST breeders can be achieved and optimized by extending the ion capture time in the trap. However, for long breeding times, charge exchange with the residual gases is a limiting factor. At the same time, the accumulation of residual-gas ions can also compensate the electron-beam space-charge and axial potentials reducing the effective current density and axial potential experienced by the trapped ions. As a result, this can induce ion losses. The development of high-current, high current-density electron guns for EBIST breeders is not only needed for efficient capture of high-rate beams, but additionally for reaching lower *A/Q* values with higher efficiencies within shorter breeding times. This would also extend the energy limit of post-accelerated beams to isotopes having shorter half-lives.

<u>Pulse width of extracted HCIB</u>

The ion pulses extracted from EBIST breeders have often an instantaneous pulse intensity (the ratio of the number of extracted ions to the pulse width) that is too high for efficient detection of all ions, or related events, due to the dead (or processing) time of detection systems. The maximum detection rate of these systems typically ranges from 0.1 to 100 MHz. The present technique consists of reducing the instantaneous rate by extending the extraction period within the existing injection-breeding-extraction timing sequence. This is performed *in situ* by slowly controlling the EBIST trapping potential to slow down the release of ions with the use of an optimized extraction potential function, and hence spread in time the width of the extracted ion pulses [LAPI2017]. The extracted pulse can be as short as 25 µs and extended to tens or hundreds of ms if required [LAPI2018]. In many cases, the extraction period has to be extended to as long as the breeding time and for high rates, can exceed a duty cycle of 50%. This can represent a problem for accelerators that have to operate in a pulsed mode to limit power. Moreover, this long extraction time can sometimes be a source of ion losses if the cooler-buncher, which accumulates continuous beams during this period, is near its space-charge capacity.

Future facilities will increase the production rates of RIB by several others of magnitudes, putting more constraints on the instantaneous pulse intensity extracted from EBIST breeders. The implementation of high current, high current-density electron guns will help shorten the optimum breeding times, and hence spread in time (over a higher number of ion pulses) the rates delivered to users. An approach that is being developed to eliminate the time lost in the injection-breeding-extraction sequence caused by a long extraction is to add a debuncher ion trap downstream of an EBIST breeder [UJIC2019]. After charge breeding, a short EBIST pulse would be injected into a linear Paul trap. There, the pulse can be stretched in time up to 1s during the same time that charge breeding simultaneously occurs in the EBIST for the next cycle. A debuncher could allow for a (quasi-)continuous beam to be delivered to users that is only interrupted for EBIST injection and extraction of less than 1 ms. The development of debuncher ion traps will be needed to efficiently spread over time high-intensity ion pulses ejected from EBIST breeders. More effort should be put in this direction.

Injection, charge-breeding & extraction simulations

Increases in production rates at RIB facilities can lead to an increase in the emittance of the beams injected into EBIST breeders. The beam emittance from a cooler-buncher operating in the pulsed mode near its charge capacity can exceed the transverse geometrical acceptance of the EBIST electron beam of 10 – 50 mm mrad. In EBIST breeders, the injected ion pulses are dynamically captured between two electrostatic potential barriers. In non-ideal cases, the beam emittance is larger than the acceptance. However, ions of large transverse velocities can still cross at times the electron beam within the capture time. Although they experience a lower "effective" electron-beam current density, they can still be ionized to high charges states, but with longer breeding times. For continuous injection, ions are injected with a kinetic energy slightly higher than a trapping barrier potential within a limited energy acceptance. The ion capture efficiency relies on their low kinetic energy in the trap and the probability that the ions be ionized to higher charge states before a trap roundtrip. As for pulsed injection, ions having large transverse velocities can yet cross the electron beam and be captured. Though limited by the EBIST geometrical efficiency and injection optics, in both modes, the effective acceptance is significantly larger than the transverse geometrical acceptance of the electron beam. The charge-state distribution and breeding efficiency in single states depends on the injected beam emittance. Moreover, the injection of intense beams can significantly neutralize the electron-beam space-charge potential. This reduces the ability of the electron-beam and axial potentials to confine ions, broadening the charge-state distribution (lower efficiencies in single charge states), inducing ion losses, and increasing the extracted beam emittance. Being able to accurately calculate the acceptance of EBIST breeders and how they perform with intense beams and near neutralization is critical to predicting breeding efficiencies and the emittances of extracted HCIB.

Over the years, several EBIST groups have developed codes to numerically simulate injection, charge breeding (charge-state distribution and ion dynamics), and extraction, all separately [PENE1991, MARG1994, KALA1998, MARR1999, LIU2005, BECK2007, LU2009, DICK2013, KITT2015, EBIS2021]. Many injection simulations estimate the transverse acceptance of EBIST breeders from particle tracking using as only trapping condition a complete overlap or a partial crossing of the injected ions with the physical size of the electron beam, ignoring the electron-impact ionization process. However, in [KITT2015], Monte-Carlo simulations including electron-impact ionization were first conducted to estimate an ReA EBIST breeder's transverse acceptance in the continuous injection mode with satisfactory agreement with measurements. Most breeding codes incorporate detailed atomic- and plasma-physics processes to simulate charge breeding and the dynamics of the trapped ions, but as initial condition, the injected ions or atoms are positioned within or near the electron beam with only assumed spatial distributions, neglecting the emittance of the injected beam. Similarly, many extraction simulations are based on assuming initial spatial and velocity distributions for the trapped ions. The space charge of the injected and residual gas ions within the electron beam, which can significantly affect the extracted beam emittance, is often neglected. FAR-TECH Inc. has recently developed a promising (commercial) particle-in-cell code (EBIS-PIC2D) to numerically simulate injection, charge breeding, and obtain detailed information for extraction simulations [ZHAO2012, ZHAO2014, JINS2015]. An electron-gun simulation code (PBGUNS) can be used to simulate the electron beam and provide solutions for its electrostatic potential and electron density when the ions are first injected. EBIS-PIC2D models the evolution of the space-charge self-consistently (electrons and ions) in 2D (radial and longitudinal coordinates), while allowing for particle motions (and tracking) in the full 3D spatial dimensions. The code models electron-ion and ion-ion collisions with injected and residual gas ions as well as the most important physical processes in EBIST. The energy spread and angular divergence of the injected beam can be defined as input parameters, but the code can also be modified to accept any injected beam profiles. It can calculate charge-state distributions as well. The position and velocity of trapped HCI ions within the electron beam can be obtained to estimate extracted beam emittances. The breeding process was benchmarked with TEST-EBIS measurements at the Brookhaven National Laboratory. Provided sufficient computing power, EBIS-PIC2D can be used to simulate injection and the charge-breeding process in all EBIST relevant dimensions. It can simulate breeding times of less than tens of ms, but for longer times, when the axial ion distribution reaches uniformity, a 1D version can be evoked to reduce the computing power. This can be a limiting factor when dealing with long breeding times and intense beams. The development and improvement of such codes is needed. This appears accessible in the future with increasingly faster computers and the availability of computing centers. It would be advantageous if such codes could include accurate 3D geometries to provide the phase-space of injected beams in the trap (in the charge-breeding region) as well as extracted beams.

*2.3.1.3. Capabilities*

Multi-charge-state acceleration

After charge breeding, HCI are extracted from EBIST breeders and accelerated to a *Q/A* separator where only a single charge state is selected for post-acceleration and delivered to experimental set-ups. All other charge states are lost which limits the breeding efficiency (up to ~30% for best cases). The *A/Q* spectra of EBIST breeders can contain few contaminants, which can be advantageous for the acceleration of multiple charge states all at once to increase efficiencies. Multi-charge-states acceleration can be accomplished with current accelerator technologies for charge states within $\Delta(Q/A)/(Q/A) \lesssim 0.04$ [FRIB2020]. This corresponds to more than 2-3 charge states for $Q \gtrsim 30$. A *Q/A* separator could be either used in a non-dispersive mode or else various charge states could be regrouped with a second magnet after the separator before transport to the post-accelerator. Prior to injecting beams into a radio-frequency (RFQ) linear accelerator, for instance, the velocity of each charge state would have to be re-adjusted with a radio-frequency cavity placed after a multi-harmonic buncher specifically designed for that purpose. Multi-charge state acceleration can be foreseen in the future and is an area of development at post-accelerator facilities that has the potential to significantly increase breeding efficiencies by a factor of more than 2, to potentially exceed ~60% for the best cases.

Manipulation of a nuclear-isomer population

As techniques for producing RIB continue to improve, opportunities arise to study nuclear reactions on nuclear isomeric states within the beams. An isomer may be of interest for probing otherwise inaccessible nuclear structure information or else astrophysical reactions on the isomer can play an important role in the reaction network. Being able to manipulate the population of the isomer to isolate it from the ground state in nuclear-reaction studies is important for state-selective measurements.

For isomers having half-lives ranging from tens of ms up to several seconds, an isomer population can be manipulated for nuclear-physics studies by varying the ion breeding time in EBIST breeders. In recent tests with $^{38}$K [CHIP2018], the population ratio of the first-excited state (0+ isomer) to the ground state was measured for different breeding times along with the corresponding charge-breeding efficiency of He-like $^{38}$K needed for post-acceleration. He-like $^{38}$K beams were subsequently delivered to an experiment using two different breeding times: 150 ms and later 1500 ms (average combined Cooler-Buncher+EBIST trapping time of ~2250 ms). This reduced the isomer-to-ground state population ratio by approximately a factor of 2. This is a capability of EBIST breeders, operating in a pulsed mode, that can expand the range of tools for the study of nuclear isomers.

*In-trap* radio-frequency cleaning

RIB injected into EBIST breeders can often be contaminated with isobars. Some isobars can be eliminated before injection with a high-resolution magnetic mass spectrometer (HRMS) or using a multi-reflection Time-of-Flight (TOF) mass spectrometer (MR-TOF), for instance. Isobars and other contaminants having the same *A/Q* as that of the rare-isotope ion species being accelerated can also originate from the EBIST residual gas. The addition of an HRMS or MR-TOF to the *Q/A* separator can potentially remove contaminants. However, to achieve high resolution, such devices typically necessitate a small beam emittance, which can be difficult with EBIST beams without introducing significantly losses. Moreover, these additional accelerator components can significantly add to the construction cost of a facility. A technique to eliminate or mitigate the presence of iso-*A/Q*'s in post-accelerated beams consists of using stripper foils, but this can deteriorate the beam properties and lead to considerable losses.

Penning-trap mass spectrometry can achieve a high mass resolving power [DILL2018]. By turning off the electron beam, EBIST can be operated in a so-called Penning-trap mode. It is a well-established technique to measure the lifetime of metastable electronic excited states of HCI in EBIST systems used for atomic spectroscopy [LAPI2005]. Certain EBIST breeders already operates with an electron-beam on/off duty cycle to prevent glow discharges and improve vacuum. Excitation of the motion of highly charged krypton ions has been tested with the Livermore (LLNL) EBIT in the Penning-trap mode for Fourier transform ion cyclotron (FTICR) mass spectrometry [BEIE1996]. With a radio-frequency (dipole) field applied to pole-tip electrodes in the trapping region, a resolving power of $\Delta(A/Q)/(A/Q) \sim 3 \times 10^{-4}$ was achieved. Based on this work, the ReA EBIST and TITAN EBIT breeders incorporate split trapping electrodes for RF ion excitation. This was successfully tested with the TITAN EBIT to clean He-like $^{40}$Ar with the presence of an electron beam of low current [LAPI2010]. These split electrodes, however, are long (~7 cm) and cover a highly inhomogeneous magnetic field that considerably limits the resolving power to less than ~0.01. The implementation of mm-long pole tips would allow for ions to experience a more homogeneous magnetic-field region resulting in a higher resolving power. RF excitation of trapped ions was studied with an EBIST at the University of Frankfurt [ZIPF2000]. Understanding how RF cleaning could be efficiently implemented in EBIST breeders would be highly beneficial to the field.

Carbon-ion (hadron) therapy

Carbon-ion cancer therapy is becoming a market-viable option for treatment of radiation-resistant tumors. The first-generation accelerator facilities devoted to hadron therapy proved the concept and established dosimetry techniques and quality assurance. The focus is now in two areas: (i) reducing the cost of the accelerator; improving its efficiency and reliability to make it more affordable and (ii) in-depth studies of the radiation damages to cells. For the former, synchrotrons are said to be high maintenance and high-cost machines, not suitable to a hospital environment. The LINAC technology is mature. Non-superconducting LINAC-based designs provide significant reduction in construction and maintenance cost. For such accelerators, the ion source has to match their injection requirements of beam intensity and repetition rate. The ion source for this application has to provide $C^{6+}$ pulses with a specific pulsed structure: 300–400 Hz repetition rate, $10^8$ ions per pulse, and ~1.5 μs pulse length. The general requirements to the ion pulse structure (short, intense, high repetition rate) make an EBIST (breeder) an ion source of choice for LINAC-based systems. For instance, EBIST systems are considered ideal for hadron therapy using iRCMS [SATO2006].

$^{11}$C has been suggested as a primary isotope for carbon-ion therapy, normally performed with the stable isotope, $^{12}$C. $^{11}$C is radioactive and decays by positron emission with a half-life of 20 min. This makes it a good candidate to verify dose delivery by positron emission tomography (PET). In conventional $^{12}$C-ion therapy, ECRIS systems are often employed as ion sources. They can produce a continuous carbon beam of high intensity that can be chopped in short pulses to meet the accelerator requirements. This reduces the total beam intensity available for treatment, which is acceptable with $^{12}$C. As it is radioactive, however, $^{11}$C can only be produced in small quantity. Although the total amount of produced $^{11}$C can be sufficient for treatment, it is generally unaffordable to have a continuous $^{11}$C beam, provided by an ECRIS breeder, sliced for injection into an accelerator. This would generate significant losses making the treatment time-consuming and, in most cases, impossible due to the low intensity of the delivered beam.

EBIST breeders do not currently meet the requirements of ion pulse intensity and repetition rate for injection into LINAC-based systems. Both parameters are mainly defined by the current and current density of the electron beam. A future EBIST breeder having a high-current, high current-density electron beam for high charge capacity and fast breeding shows potential to meet the demands for carbon-ion therapy with LINAC-based accelerators [SHOR2016, MERT2017].

### *2.3.2. ECRIS breeders*

*2.3.2.1. Design and operation*

In ECRIS breeders, multiply charged ions are produced with an electron-ion plasma confined within axial and radial magnetic-field minima created by non-superconducting solenoids and permanent magnet multipoles. The plasma is generated by injection of microwave in a plasma chamber, where electrons from a support gas (e.g., He, $O_2$, $N_2$) are produced by ionization and resonantly excited. The resulting broad electron-energy distribution leads to a broader ion charge-state distributions compared with EBIST breeders. A schematic layout of an ECRIS charge breeder is presented in Fig. 2.3.3. ECRIS breeders mostly operate in a mode of continuous injection and extraction. Injected ions of tens of keV in energy are injected into one side. The ions are electrostatically decelerated upon injection, and by Coulomb collisions in the plasma are thermalized and deflected to the core of the plasma region. There, they are accumulated and confined by the potential dip where they undergo stepwise (mostly single) ionization via collisions with electrons. As injection and charge breeding occur, HCI are simultaneously extracted.

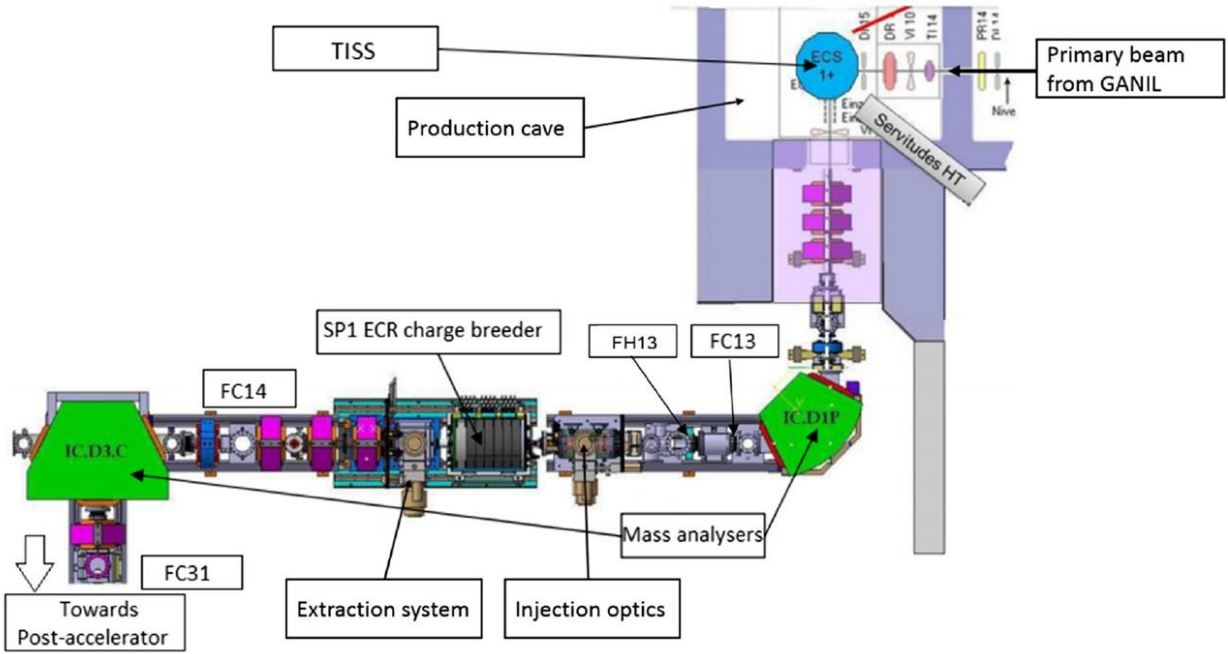

*Figure 2.3.3. Schematic layout of the ECRIS breeder at the SPIRAL1 facility [ANNA2021].*

ECRIS breeders are built for high breeding efficiencies and maximize production of high charge states. The large number of plasma electrons provides a high electron density of less than $5 \times 10^{11}$ cm$^{-3}$ with an electron temperature in the range of 1-2 keV. They have a high charge capacity of typically less than about $10^{14}$ elementary charges, best for capture of beams of high rates. They typically have lower electron densities than EBIST breeders resulting in longer breeding times. Ions are injected into the plasma with a low kinetic energy carefully adjusted to reach and be efficiently trapped by the few-eV potential dip. Depending on the mass of the elements, the extracted *A/Q r*ange is typically 2 – 7. The efficiency for ECRIS breeders is the 10–15% range. The overall parameters/characteristics of ECRIS breeders in operation, under construction, or being planned (ANL [VOND2008], KEK [IMAI2008], LPSC [GELL2006], RAON [WOO2015], SPES [PRET2009], SPIRAL-GANIL [DELA2012], Texas A&M [TABA2019], TRIUMF [AMES2008]) are listed in Table 2.

*Table 2.3.2. Typical operational parameters of ECRIS breeders.*

| A/Q range | ≲ 7 |
|---|---|
| Typical mass range | ≲ 200 |
| Breeding efficiency | ≲ 10-20% |
| Beam emittance (1x RMS) | ≲ 0.1 mm mrad |
| ΔE (energy spread) | ≲ 50 eV |
| Electron temperature | ~ 1-2 keV |
| Electron density | ≲ $5 \times 10^{11}$ cm$^{-3}$ |
| Capacity (units of elementary charge) | ≲ $10^{14}$ |
| Microware frequency | ~10-14 GHz |
| Max. magnetic field | ≲ 1.5 T |
| Trapping volume | ≲ 300 cm$^3$ |

*2.3.2.2. Limitations and areas of development*

<u>Contamination</u>

High-energy ions and electrons (as well as x-rays) can escape from the ECRIS breeders' plasma and interact with the plasma chamber's wall and surrounding electrodes. This can generate a pervasive ion background substantially contaminating extracted ion beams limiting the performance of ECRIS breeders. Considerable effort has been spent over the years to mitigate this background. Careful cleaning of the plasma chamber and surrounding electrodes is mandatory (ultra-high vacuum (UHV) and $CO_2$ cleaning [VOND2015]). The use of UHV equipment as well as ultra-pure support gases, construction materials, liners, and surface coating have been introduced to limit the amount and type of impurity. For instance, the INFN-LNL, LPSC, and SPIRAL-GANIL have collaborated to address the issue by testing plasma chamber liners of different mono-isotopic material such as pure tantalum or niobium [MAUN2021]. Pure construction materials made of heavy elements to limit the sputtering yield is being investigated. A cryogenic plasma chamber has also been proposed [MAUN2022]. The relative intensity of the background of ECRIS breeders may become less of an issue at future facilities producing RIB of high rates. Nevertheless, the development of new techniques and design to reduce the background level of ECRIS breeders is primordial.

<u>Fine-frequency tuning and two-frequency heating</u>

Post-accelerators can accept a large array of *A/Q*'s for delivering a wide energy range (from a few MeV/u up to tens of MeV/u). ECRIS breeders generally operate with a single and fixed heating microwave frequency leading to a fixed charge-state distribution given for each element. This constrains the energy range of the post-accelerated beams delivered to experimental set-ups and users. Slight changes in the microwave heating frequency of the plasma or the addition of a second heating frequency can affect the plasma conditions and lead to substantial changes in the charge-state distributions [TRAS2003]. In particular, the use of two-frequency heating was demonstrated with the ECRIS breeder at the Argonne National Laboratory where a 2-fold increase was observed for high charge states of $^{129}$Xe (25+, 26+, and 27+) [VOND2012]. Unlike EBIST breeders, as ECRIS breeders are operating in a continuous mode, optimizing specific charge states can be difficult for fixed plasma conditions. Fine-frequency tuning (that can be combined with a variable magnetic field) and two-frequency heating can allow for the optimization of low, medium, and high charge states, while conserving the total charge-breeding efficiency. This can help stabilize the plasma and find a compromise between efficiency and breeding time, particularly needed for rare-isotopes of short half-lives. For instance, this would allow for low charge states (short breeding times) to be optimized to maximize the breeding efficiency of short-lived isotopes, extending post-acceleration with ECRIS breeders towards elements further away from the valley of stability. R&D is ongoing and needed in the field to optimize and manipulate the range of charge states (and isotopes) available from ECRIS breeders. This should be prioritized.

Injection, charge-breeding & extraction simulations

The charge-breeding efficiency of ECRIS breeders has improved over the years, but generally remains by a factor of 1.5 to 2 lower than that of EBIST breeders. The ion injection efficiency into the plasma can considerably affect the overall efficiency of the device. Injected ions are electrostatically decelerated to a suitable kinetic energy and by Coulomb collisions with plasma ions are thermalized. They then reach the plasma core, where they undergo ionization with plasma electrons and become confined by the potential dip. To reach high breeding efficiencies, the kinetic energy of the injected beam as well as its energy spread and emittance are critical as the energy (~a few eV) and transverse acceptance for ion capture by the plasma are small. The use of a cooler-buncher has being proposed prior to injection into the SPIRAL1 ECRIS breeders to improve the beam properties of injected beams [ANNA2021]. Injection is a complex process that is sensitive to plasma parameters (e.g., type of plasma ions, plasma density, etc.). As they have low kinetic energy upon injection, the trajectory of the injected ions can also be particularly sensitive to magnetic-field asymmetries [VOND2012]. The injection optics has to be carefully designed. Charge-bred ions are extracted and accelerated from ECRIS breeders by applying a potential difference between an extraction electrode, which is part of the plasma region, and an adjacent puller electrode. The shape of the plasma boundary (meniscus) at the extraction electrode is space-charge dependent, and varies with plasma parameters. The curvature of the meniscus can strongly affect the profile of the extracted beams [TRAS2003]. Depending on its shape, the distance between the extraction electrode and puller as well as the design of the extraction system have to be adjusted to optimize beam quality. A better understanding of the space-charge force on the extracted ions would help in the design of extraction systems.

Over the last years, progress in numerical simulation tools (e.g., Monte-Carlo, particles in cells, 3-dimensional (3D) electromagnetic solvers, etc.) and advances in computing power have allowed for a better understanding of the injection process into ECRIS breeders and the overall plasma. A recent 3D numerical simulation code developed by the INFN ion source group including electron-impact ionization, Coulomb collisions, and other collisional processes as well as a self-consistent description of the plasma is showing promising results in the direction of accurately modelling ECRIS breeders [GALA2016a, GALA2016b, TARV2016, GALA2019, GALA2020]. This code reproduces well important aspects of the device such as injection efficiencies and how slight changes in the microwave heating frequency can affect the plasma density. Important information on injection was obtained from the simulations. For instance, the first ionization steps of the injected ions are fundamental to the capture process and the number of injected ions overcoming the maximum of the magnetic field at the injection side plays a role than is more important than the ion capture process in the plasma. This explains the narrow injection energy acceptance. Most importantly, the code demonstrates the predictive power of current numeral simulations. The development of advanced codes or combination of codes [ANNA2021] simulating injection, charge breeding, charge-state distributions, and extraction, including the ECRIS plasma, are necessary for an improved understanding of their sensitivity to plasma parameters, electric and magnetic forces, and source parameters. In this respect, a better understanding of the stopping power of ions in highly ionized matter would be beneficial. Effects that have been of interest over the years to increase the production of higher charge states, improve efficiencies and damp instabilities such as fine-frequency tuning, two-frequency heating, biased disks enhancement could be further investigated with such codes. In parallel, experimental studies to support, serve as input parameters, or validate the result of those simulations should continue.

Afterglow effect

The afterglow effect, observed by turning off the microwave power, produces intense short ion pulses (200 µs -10 ms) that can lead to a 10-fold increase in the intensity of HCIB [TRAS2003]. High charge states experience a higher trapping potential in the plasma core than singly charged ions, for instance. During continuous extraction, the range of charge states confined within the plasma is higher than those released because only ions in the high-energy tail of their energy distribution can escape the plasma potential. After turning off the microwave power, however, the plasma electrons leave, removing the confining potential. This process is comparable to lowering the electrostatic potential barrier of an EBIST breeder to release all trapped ions. ECRIS breeders were built for mainly operating in a mode of continuous beam injection and extraction. The advantage of these devices is their high charge capacity, allowing for capture of beams of high rates that are too intense to be efficiently confined in a cooler-buncher and/or the electron-beam space-charge potential of an EBIST breeder. An ECRIS breeder extracting ions in the afterglow mode can be a promising option for efficient production of high charge states of intense beams. As ion injection is continuous, a short duty cycle of the turn-off time of the microwave power would be needed to limit the loss of ions that cannot be injected during the extraction time. However, for experiments that are unable to accept short intense pulses, the addition of a downstream debuncher could allow for a broad range of experiments requiring stretched ion pulses to limit the instantaneous rate. The use of the afterglow mode in charge breeding can lead to a better control of a charge-state distribution by varying the afterglow on-off repetition rate. This would be advantageous for maximizing the charge-breeding efficiency of short-lived isotopes. The afterglow effect and its reliability during operation should further be studied and improved.

### Towards lower *A/Q*'s

Several development areas have been mentioned above to produce higher charge states in conventional (non-superconducting) ECRIS breeders that would extend the energy range of post-accelerators to accelerate isotopes of shorter half-lives to higher energies. The production of high charge states is limited in the most part by the ionization rate with respect to the rate of charge exchange with neutral gases. Reducing the flux of neutral gas needed for plasma production may help, but can also lower the electron density. Increasing the electron density in ECRIS breeders is key to improving the ionization rate for higher charge states. Higher magnetic confinement can increase the electron density and hence the maximum charge state in a distribution ($q_{max}$), as described by Geller's empirical scaling law $q_{max} \propto \log B_{max}$, where $B_{max}$ is the maximum of the magnetic mirror field [GELL1990a, GELL1990b]. All ECRIS breeders use resistive and/or permanent magnets and make use of soft-iron yokes, for instance, to further increase the plasma-confining magnetic field. The development of new designs in this direction may be beneficial. An increase of the magnetic field allows for the use of a higher heating microwave frequency, $\omega$ From Geller's scaling laws, the maximum charge state significantly increases with the microwave frequency as $q_{max} \propto \log \omega^{7/2}$, while the extracted beam intensity scales as $I_q \propto \omega^2$. For the production of increasingly higher charge states, the implementation of superconducting magnets on future ECRIS breeders is being discussed at the Laboratoire de Physique Subatomique et Cosmologie (LPSC). This technology is already being used with ECRIS injector at accelerator facilities. The implementation of superconducting ECRIS breeders would be more cost-efficient than adding accelerating structures for the same accelerated beam energy.

## 2.4. Machine Learning (ML) and Artificial Intelligence (AI) for improved ion source performance and more efficient long-term operations (Lead Contributors: **Todd, Benitez**)

Regardless of ion source type, the production of highly charged heavy ion beams typically involves the optimization of tens of control parameters. Optimization of these parameters leads to better performance and higher production. As plasma physics is notoriously complex and demanding of computer simulation, the plasmas at the hearts of highly charged ion sources typically do not have computer simulations to guide their operational optimization and the advances in performance have come almost exclusively from explorations of the operational space.

Currently, most control of these sources is human-driven. The optimization of these multi-parameter spaces is slow and can be both difficult and time-consuming for humans to navigate. An example of this can be seen in the ECR ion sources. The two best-performing sources, VENUS at LBNL and SECRAL-II at IMP in Lanzhou, China, have approximately three decades of tuning experience between them, yet every year they still produce record beams. These record beams are primarily a result of additional searches of the vast operation space of these sources. The vastness of this operation space and the tediousness of these searches makes them ripe for Machine Learning where source operation is computer-driven and source diagnostics are fed back into the program in order to determine operation modes that improve performance. This can take on many forms: in the simplest mode a single parameter is maximized such as the current of a specific ion species. However, by monitoring multiple diagnostics while exploring the operation space, Machine Learning allows the possibility to optimize solutions that are a compromise between increased beam current and reduced beam emittance, say, for better overall accelerator transmission. These optimizations would be difficult, if not impossible, for humans to perform on reasonable time scales.

In addition to high performance, sources attached to accelerators are often tasked with delivering stable ion beam currents for weeks at a time. These long periods of operation leave the sources subject to drifts of control parameters. Typically, these drifts are countered by a human operator once performance has deteriorated to the point that it is noticed by a human or where it has affected accelerator performance. Often these countermeasures require interruptions in beam delivery while changes are made to the source. Machine Learning has the potential of greatly improving long-term source stability as a computer program can continually monitor source conditions and, by using learning on historic data, both identify early signs of impending instability and provide countermeasures to prevent drifts that lead to losses of operation time.

Machine learning is a young, but vigorous, area of research for both accelerators and ion sources. Controlling and optimizing the tens of parameters is going to be difficult, and as the application of machine learning to the task is relatively recent, there are few successes of note in the literature that are directly related to highly charged ion sources. However, recent results demonstrating the control and optimization of a tokamak plasma via machine learning [DEGR2022] clearly demonstrate that this is not an impossible task and that it should be expected that real successes applying machine learning to highly charged ion source plasmas will come in the not-too-distant future.

## 3. Future development priorities

### 3.1. Electron Cyclotron Resonance Ion Source

As discussed in the aforementioned section that the production of intense HCIB scales with higher magnetic field and operation at higher heating frequency. The progress made between the 1$^{st}$ generation ECRIS and current 3$^{rd}$ generation ECRIS such as VENUS and SECRAL have shown that Geller's scaling law predicts the trend of ECRIS advancement and what can be expected from a 4$^{th}$ generation and future generations of ECRIS. The critical challenge in advancing the state of ECRIS lies in achieving higher magnetic field to support operation at higher heating frequency and to provide stronger plasma confinement. The future development of ECRIS should go forward:

- Development of a 4$^{th}$ generation ECRIS. The R&D work is to develop a magnet system that can produce the needed magnetic minimum-*B* field configuration with either NbTi or Nb$_3$Sn conductor. This is the most urgent task that needs sufficient funding to move forward right now.
- Furthermore, if the MARS magnet geometry gets validated in ECRIS, an HTS or a hybrid (HTS + other conductors) based MARS magnet capable of generating magnetic fields for operating at ~ 80 GHz will be the pathway to a future generation super-performing ECRIS.

After the issue of higher field magnets has been overcome, higher frequency microwave generation with suitable power level in cw mode for ECRIS operations could be another important task and that should be addressed in the future, too.

### 3.2. Laser Ion Source

The laser ion source is a simple ion source that can provide intense pulsed beams. However, there are some limitations at present.

1. High repetition rate operation. The laser power density must be increased to obtain high charge states, but these results in cratering of the target, and the target supply cannot keep up during high repetition rate operation. To solve this problem, tape targets can be prepared and wound at high speed, or liquid targets can be prepared.

2. Long pulse widths. The pulse width depends on the velocity distribution of ions in the generated plasma. If the velocity at which the center of gravity of the plasma moves can be reduced, the time difference within the pulse of the plasma reaching the extraction field can be increased. To achieve this, a foil with a thickness of less than 1 μm could be used as a target. However, in this case, the plasma would not have enough initial velocity and would be expected to diffuse in all directions. To prevent this, ionization in a high magnetic field and limiting the direction of diffusion may prevent the plasma density from decreasing.

3. Generation of high charge-state ions from heavy elements. It is useful to use a stable solid-state laser to generate stable plasma. Most solid lasers on the market have pulse widths of 5 to 10 ns, but the heating time is too long to obtain high charge states. It is necessary to optimize the laser irradiation conditions for the species and charge state. The laser pulse width survey would be a useful strategy to have optimum laser irradiation conditions.

4. Instability of the ion beam is often caused by the roughness of the target surface. To prevent this, the beam can be further stabilized by adjusting the lens position while precisely measuring the position of the target and focus lens.

The above list will be the subject of R&D in the near future.

### 3.3. Charge breeders

Future RIB facilities are expected to significantly increase the production rates of rare isotopes by several orders of magnitude. Such facilities may necessitate the implementation of the two systems, EBIST and ECRIS breeders, for complementarity in charge breeding RIB of different intensity ranges (ECRIS: high-intensity, EBIST: low-intensity). Effort to improve their performance both in parallel is therefore important.

R&D in high current, high current-density electron guns for EBIST breeders is needed for efficient charge breeding of high-intensity beams. This would enable two operating modes: 1) pulsed injection at high repetition rates to remain within the charge-capacity limit of the upstream cooler-buncher and 2) continuous injection with longer breeding times for beam rates exceeding the buncher's capacity. Such guns would also allow for the reach of lower *A/Q* values with higher efficiencies with shorter breeding times, extending the energy limit of post-accelerated beams to isotopes having shorter half-lives. A long-term target goal of 4 A and ~5500 A/cm$^2$ is suggested.

The pervasive background of ECRIS breeders can substantially contaminate RIB of low intensity. Significant effort has been spent over the years to mitigate this background. Although, the relative intensity of this background may become less of an issue with RIB of high rate, the development of new techniques and designs to reduce it should be considered a priority.

ECRIS breeders generally operate with a single and fixed microwave heating frequency leading to a fixed element-dependent charge-state distribution. This limits the energy range of the post-accelerated beams. Fine-frequency tuning and two-frequency heating can help stabilize the plasma and modify charge-state distributions to optimize low, medium, or high charge states. Along the same line, the afterglow effect produces intense short ion pulses that can lead to a 10-fold increase in intensity of high charge states. The afterglow mode could also be used to shift charge-state distributions by varying the afterglow on-off repetition rate. This would be advantageous for maximizing the charge-breeding efficiency of short-lived isotopes. Fine-frequency tuning, two-frequency heating, and the afterglow effect can allow for low charge states (short breeding times) to be optimized to maximize the breeding efficiency of short-lived isotopes, extending with these isotopes the energy limit of post-accelerated beams from ECRIS breeders. R&D is ongoing and needed in this direction.

Constant progress in numerical simulation tools and advances in computing power is enabling the development of increasingly detailed self-consistent numerical simulations codes for ECRIS and EBIST breeders. A promising ECRIS-breeder 3D code has recently allowed for a better understanding of injection and the ECR plasma in agreement with observations, which demonstrates the predictive power of such simulations. The development of (complete) 3D codes including all aspects of ECRIS-breeder to simulate injection, charge breeding, charge-state distributions, and extraction (including the plasma) is needed to understand their sensitivity to source parameters. Similarly, a particle-in-cell code (EBIS-PIC2D) have been developed and is commercially available to simulate injection, the charge-breeding process, and provide initial conditions for ion extraction simulations. The 2D version of the code is limited to short breeding times, requirement less computing power. The development and improvement of such codes is needed to deal with long breeding times and intense beams, which leads to high neutralization of the electron beam. It would be beneficial if such codes would include accurate 3D geometries to provide the phase-space of the injected beams in the trap and that of the extracted HCIB. Overall, the development of *complete* EBIST-and ECRIS-breeder 3D numerical simulation codes (full 3D geometries) is highly necessary to improve their performance (increase efficiencies and provide beams of higher quality) and should be set as a (continuous) long-term goal. This appears accessible in the future with increasingly faster computers and the availability of computing centers.

### 3.4. Machine learning and Ion sources

The optimization of an ion source's operation space, some tens of parameters, is difficult and time-consuming for a human.  The application of machine learning to this task promises advancement in both source performance and the maintenance of stability.  The research into the optimization of ion sources via machine learning is still a young area of research, but the recent successes controlling and optimizing a tokamak plasma via machine learning [DEGR2022] demonstrates that the same should be possible for ion sources.  It should be expected that efforts directed at the application of machine learning to ion source operation would result in real successes in the not-too-distant future.

## 4. Conclusions

Ion sources producing intense HCIB play an indispensable role at heavy ion accelerator facilities worldwide for particle- and nuclear-physics research. As the complexity of microelectronics grows, they are also crucial to test radiation effects at high beam energy. With present and upcoming accelerator upgrades and advances in accelerator technologies, the demand for high-quality high-intensity HCIB is increasing. Significant R&D is now becoming essential to refine our understanding these ion sources needed to improve their performance in order to match the requirements of future accelerator facilities. Constant progress in computing power and tools for advanced numerical simulations as well as in Machine Learning is persistently opening the door to new R&D opportunities.

EBIST or ECRIS are presently the only ion sources employed as charge breeders of RIB. In the future, however, novel ion sources currently in used as injector may eventually make their way to RIB charge breeding, such as Electron-String Ion Sources [BOYT2019] and a novel EBIS-type trap, named ITRIP, being studied to boost HCIB of high intensity to higher charge states starting from an ECRIS injection [HUAN2021].

R&D in ion sources for production of high-intensity HCIB can also lead to breakthroughs in other fields such as accelerator mass spectrometry (AMS) using a high-current ECRIS [WU2020] or medical applications. For instance, the development of high-current, high current-density electron guns for EBIST breeders is of interest for hadron therapy [SHOR2016, MERT2017].

A LIS can provide high current, high brightness beams with a wide variety of charge states from many species. The structure of the LIS is simple and the illumination chamber is easily isolated electrically. Therefore, a massive safety fence can be eliminated. In BNL, we are regularly operating both a low-charge state source and a high-charge-state source with DPIS with good stabilities. Solenoid plasma manipulation was established and an LIS has improved flexibility. Mentioning the application, a LIS is suitable for neutron production using inverse kinematics because it can produce a high intensity pulsed ion beam. Furthermore, by achieving a high repetition rate, LIS has been proposed for use in the purification of medical isotopes. It is also promising as a heavy ion generator for accelerator-based inertial fusion.

**Acknowledgements**

The authors thank E. Beebe, G. Machicoane, L. Maunoury, B. Roeder, D. May, G. Tabacaru, F. Wenander, H. Pahl, R. Vondrasek, Y.-H. Park, and B. Schultz for discussions and providing valuable information.